\documentclass[aps,prb,twocolumn,nopacs,amssymb]{revtex4}

\usepackage{graphicx}
\usepackage{dcolumn}
\usepackage{bm}
\usepackage{color}

\usepackage{amsmath}
\usepackage{amssymb}

\def\be{\begin{equation}}
\def\en{\end{equation}}
\def\bea{\begin{eqnarray}}
\def\ena{\end{eqnarray}}
\def\bec{\begin{equation}\begin{array}{rcl}}
\def\p{\partial}

\def\gs{\gtrsim}
\def\ls{\lesssim}

\def\ab{{ij}}
\def\bc{{jk}}
\newcommand{\av}[1]{\langle{#1}\rangle}

\newcommand{\bi}[1]{\mbox{\boldmath$#1$}}

\begin{document}
\title{Theory of  nonionic hydrophobic solutes 
in mixture  solvent: Solvent-mediated interaction and 
solute-induced phase separation   }
\author{Ryuichi Okamoto$^a$ and Akira Onuki$^{b}$ }
\address{
$^a$ Research Institute for Interdisciplinary Science, Okayama University, Okayama 700-8530, Japan \\
$^b$ Department of Physics, Kyoto University, Kyoto 606-8502,
Japan\\
}


\date{\today}

\begin{abstract}
We present a theory of  nonionic  solutes 
 in a  mixture solvent composed of water-like and alcohol-like 
species.  First, we show relationship among  
 the  solvation chemical potential, the   partial volumes $v_i$, 
 the Kirkwood-Buff integrals,  the second osmotic virial 
coefficient, and the Gibbs transfer free energy. 
We examine  how the solute density $n_3$ is coupled to  
the solvent densities $n_1$ and $n_2$ in thermodynamics. 
In the limit of small compressibility, we show that  
the space-filling condition $\sum_i v_i n_i=1$ nearly 
holds for  inhomogeneous densities  $n_i$, where 
the concentration fluctuations of the solvent 
can give rise to a large  solute-solute attractive interaction. 
We also  derive a   solute spinodal density  $n_3^{\rm spi}$ 
 for solute-induced instability.
  Next, we examine   gas-liquid and liquid-liquid 
phase transitions induced by  a small amount of a 
  solute using    the  Mansoori, 
Carnahan, Starling, and Leland model for hard-sphere  mixtures 
$[${  J. Chem. Phys.} {\bf 54}, 1523 (1971)$]$.
Here, we assume that   
 the  solvent is close to its gas-liquid coexistence 
and  the solute interacts repulsively with the 
water-like  species but  attractively with the alcohol-like  one. 
We calculate  the binodal and  spinodal curves 
in the phase diagrams  and examine nucleation 
  for these two  phase  transitions. 
\end{abstract}


\maketitle


\section{Introduction}

Long-standing  research  has 
 been made on the role of hydrotropes 
in aqueous mixtures. Short-chain alcohols 
(such as methanol or tertiary butyl alcohol (TBA)) 
 are typical examples of  nonionic hydrotropes, which 
 have  an amphiphilic character 
but form  no ordered structures in water 
due to their  small sizes\cite{Kaler,Zembreview}.
A hydrotrope  interacts with both water and hydrophobic solutes 
attractively, so it can  improve  the solute solubility 
 as a cosolvent. 
As a unique effect, microemulsion-like    droplets  
with radii of order   $10^2-10^3$ nm emerge 
in ternary   mixtures of 
  water,    alcohol, and  a hydrophobic solute
\cite{Rus,Gri,Smi1,Zemb,Zemb1,Zemb2,Wegdam,Katz,Hori,Hou,Ani1,Ani2,Ani3}  
 as well as macroscopic domains,    
 depending on the solute   and  alcohol fractions. The former 
 have well-defined interfaces 
yielding  the Porod tail in the scattering amplitude\cite{Gri}.
This phenomenon is  known by the name of {\it Ouzo effect}\cite{Katz}. 
 In such phase separation,  a  hydrophobic solute  is 
 accompanied by  alcohol (as they  go out of the original liquid)  
and the surface tension is considerably 
decreased due to  interfacial adsorption of 
alcohol\cite{Wegdam,Zemb1,Zembreview,Hou,Ani3}. 
Furthermore, scattering experiments\cite{Zemb,Zemb1,Zemb2,Hori}  
have indicated the presence of  nanometer-sized,   
 micelle-like aggregates\cite{Tan}  in a wider {\it pre-Ouzo} 
 region  outside the binodal. 
In molecular dynamics simulation 
on water-ethanol-octanol mixtures\cite{Hori,Zemb2}, 
a large fraction of  octanol molecules  
 aggregate to form the cores  of 
 such nanoclusters, where  a large number of  ethanol  molecules 
cover and penetrate these  clusters.

In this paper,  we theoretically treat a ternary   mixture 
of a  water-like solvent, an   alcohol-like cosolvent, 
and a nonionic hydrophobic solute.  We need to properly 
account for the steric and attractive  interactions   
between the solute and  the two  solvent species 
  at very small solute fractions. To this end,  we 
use  a continuum model  of  hard-sphere  mixtures by 
 Mansoori, Carnahan, Starling, and Leland  (MCSL)\cite{Man}. 
which   is a generalization of 
the Carnahan and  Starling  model of  one-component hard-spheres  
 (CS)\cite{Car0}. For simplicity, we 
assume  attractive interactions  in  the 
van der Waals form, so we do not treat the hydrogen bonding   
among constitent protons and oxygens. 
We also  do not include  a  surface-active  character 
 of the cosolvent, which leads  to a reduction of the surface tension.  
In  this scheme, 
 we can examine  how the selective  solvation depends on 
the hard-sphere  diameters, the attractive interaction parameters, 
and the compositions.

Our model solvent is assumed to be  close 
to its gas-liquid coexistence, which is the case 
for  water-alcohol mixtures. We use   solutes which  
interact  with the two solvent  species 
 very differently. In this situation, 
 we encounter solute-induced gas-liquid and liquid-liquid 
phase transitions.  In the former, 
the  expelled solute particles form a  gas at low pressures 
in the absence of appreciable  solute-solute  attractive interaction.  
The latter  occurs  when the attractive interaction 
between the solute and the second species is sufficiently strong.  
We examine two-phase coexistence,  metastability,  and instability 
for  these  phase transitions. 
 We argue that the nucleation rate for  the liquid-liquid transition 
is much larger than that  for  
the gas-liquid one in the bulk   in  mixtures of  water, 
alcohol, and a hydrophobic solute.

As a mysterious  phenomenon, 
  long-lived mesoscopic  heterogeneities 
  (presumably phase-separated domains)   have been detected 
at very small fractions  by dynamic light scattering, which  
emerge  with  addition of a small amount of a salt, a polymer, or  
a hydrophobic compound  in one-phase states of 
aqueous mixtures or polymer 
solutions\cite{S1,S2,S3,S5,S6,S64,S7,S61,Ani1,S62}.  
From their   diffusion constants,  
their  sizes were  in the range   $10^2-10^3$ nm. 
Such   precipitation  
occurs     for various combinations of a solute  and a 
mixture solvent, so it  should generally   originate  from 
  selective solvation  of  
a   solute\cite{Bu,Oka}.

As a well-known    solute-induced gas-liquid transition, 
we   mention formation  of nanobubbles   
with dissolution  of gases such as O$_2$, H$_2$, and Ar  in 
ambient water\cite{Attard,Seddon,Uchida,Ohgaki,nanoEXP,Loshe}. 
 For example,  Ohgaki {\it et al.}\cite{Ohgaki} 
realized    bubbles of such gases  with
radius $ 50$ nm  and  
 volume fraction $ 0.01$ in quasi-steady states, 
where  bubble coalescence was   suppressed by salts added.   
Such bulk nanobubbles are usually 
 produced by breakup of large bubbles, 
while surface  nonobubbles on hydrophobic walls 
can appear  via heterogeneous nucleation. 
In this problem, it is crucial that ambient water 
 is very close to its gas-liquid coexistence. 
Then,  we can explain the bubble stability 
by including   the Gibbs transfer free energy  in the 
bubble free energy provided that  
the solute-water  attractive interaction 
is weaker than that among the water molecules\cite{nano,OkaD}. 

In the literature\cite{WidomH,Amo1,Chandler,M1,M2,M3}, 
much attention has also  been paid to 
assembly of   strongly hydrophobic  particles 
 in ambient water, where the proximity 
to  gas-liquid coexistence is  crucial\cite{Chandler}. 
In our viewpoint, it  can  be treated  as  phase separation, 
since    associated  clusters  should  
  grow if their sizes exceed a critical size. 
 However, this phenomenon  has been studied only at its inception.

Before discussing phase separation,   we present a  statistical-mechanical 
theory of  solvation  in  ternary  mixtures. 
We relate   the solvation chemical potential 
 to various physical quantities including 
 the  partial volumes $v_i$ ($i=1,2,3)$.  
For  inhomogeneous densities $n_i$, 
we discuss how the space-filling condition $\sum_i v_i n_i =1$  
is nearly  satisfied at long wavelengths 
in the limit of  small compressibility. 
Namely,  in nearly incompressible fluid mixtures, 
 the deviations  of $\sum_i v_i n_i$ from 1 
should be small\cite{Flory,Onukibook}, while the concentration 
fluctuations can be enhanced due to molecular clustering.  
This is  the case for water-alcohol mixtures.
 Then,  the solvent-mediated,  
solute-solute interaction  arises mostly from the  solute-concentration 
 coupling   for not small cosolvent fractions.

The organization of this paper is as follows. In Sec. II, we
will  present the theoretical background  of ternary mixtures. 
In Sec.III,  we will use the MCSL model 
to investigate  the solvation effects. In Sec.IV, 
we will examine   the  phase separation. 
 Additionally, we will summarize the  
theory of the partial volumes and the Kirkwood-Buff integrals 
in Appendix A, 
examine the solute-induced gas-liquid transition in a one-component solvent 
 in Appendix B, 
and present  details 
of the CS model in   Appendix C 
and MCSL  model  in   Appendix D.

\section{Dilute solute  in mixture  solvent }

We  consider nonionic ternary fluid mixtures. 
As a mixture solvent, we consider a  water-like species 
(called water) ($i=1$) and  a cosolvent $(i=2$).  
We then add a dilute  solute ($i=3$).  Their   densities  
are written as $n_i$. 
The solvent composition (cosolvent molar fraction) 
is written as 
\be 
X= n_2/(n_1+ n_2). 
\en 
The binary mixture of the solvent  
is assumed to be in one-phase states  away  from the gas-liquid  criticality. 
The   electrostatic and amphiphilic 
interactions  are not treated explicitly.  
 The  boundary effect is beyond the scope of this paper.  
 In all the calculations to follow, 
we fix the  temperature 
$T$  at $300$ K,  so we do not 
write  the $T$-dependence of the physical quantities.

{\subsection{ Solvation chemical potential}}

The Helmholtz free energy density $f(n_1, n_2, n_3)$ 
is expanded with respect to $n_3$ 
up to  second order  as\cite{nano,OkaD} 
\be 
f = f_{\rm m} +  
k_B T   [\ln (n_3 \lambda_3^3) -1 
+  \nu_{3} ]n_3 
+\frac{1}{2}  U_{33} n_3^2 , 
\en 
where   $f_{\rm m}(n_{1},n_2 )$ is the solvent free energy density 
 and  $\lambda_3(\propto T^{-1/2}$) 
is   the solute thermal de Broglie length. The 
 $k_BT \nu_{3}(n_1, n_2)$ is the  solvation 
chemical potential for a  solute particle, which   
 arises   from the interactions  with its 
 surrounding solvent.
The last term represents 
 the (direct) solute-solute interaction defined by  
\be 
U_{33} = \lim_{n_3\to 0}[({\p\mu_3}/{\p n_3})_{T, n_1, n_2}-  {k_BT}/{n_3} ],
\en  
where $n_1$ and $n_2$ are fixed and the ideal gas 
part  $(\propto n_3^{-1}$) is subtracted. 
The  chemical potentials $\mu_i =\p f/\p n_i$ and 
 the pressure $p=  \sum_{{j}} n_j \mu_j-f $  are expanded 
 up to first order corrections  as   
\bea 
\mu_i &=& \mu_{{\rm m}i}  +  k_BT \nu_{3i}  n_3 \quad 
(i=1,2),\\
\mu_3 &=&  k_B T  [\ln (n_3\lambda_3^3) +  
\nu_{3} ] + U_{33}  n_3 ,\\
p&=& p_{\rm m}  + k_B T ( 1+ \zeta_{3}) n_3 . 
\ena 
Here, we define   $\mu_{{\rm m}i}(n_1,n_2)= \p f_{\rm m}/\p n_i $ 
 and   $p_{\rm m}(n_1,n_2)=\mu_{{\rm m}1}n_1+\mu_{{\rm m}2}n_2-f_{\rm m} $.  We also introduce    
\bea 
&&\hspace{-5mm}
\nu_{31}= (\p  \nu_{3}/\p n_1)_{T, n_2}, \quad \nu_{32}= 
(\p  \nu_{3}/\p n_2)_{T, n_1},
\\
&&\hspace{-5mm}
\zeta_3= n_1 \nu_{31}+n_2\nu_{32}= n(\p \nu_3/\p n)_{T,X},  
\ena 
where   $n=n_1+n_2$ is the solvent number density. 
In this paper, we treat nearly incompressible solvents with 
small compressibility $\kappa_{\rm m}$ 
such that the inequality  
$n k_BT \kappa_{\rm m} \ll 1$ holds for any $X$. 
In terms of $p_{\rm m}$,  $\kappa_{\rm m}$ is expressed as   
\be 
\kappa_{\rm m}^{-1}=n(\p p_m/\p n)_{T,X},
\en  
where $T$ and $X$ are  fixed in the derivative. See Eq.(A3) 
in Appendix A for another  definition of the compressibility 
in many-component fluids. For example,  
$\kappa_{\rm m}= 4.5\times  10^{-4}/$MPa $\sim 0.05 /n_r  k_BT$ 
for  ambient  liquid water with density $n_r=33/$nm$^3$ at 300 K and 1 atm.

It is convenient to introduce the partial volumes\cite{Moore,Wood,OC1} 
$v_i$ for the three species in the dilute limit 
$n_3\to 0$. As will be shown in  Appendix A, they are expressed as  
\bea 
v_i &=& \kappa_{\rm m}({\p p_{\rm m}}/{\p n_i}) \quad (i=1,2),\\ 
{ v}_{3} &=& k_BT \kappa_{\rm m} ( 1+ \zeta_3 ) .
\ena 
These volumes  depend on $T$, $p$, and $X$.  For not small solutes  with 
 $ v_3\gs n^{-1}$, we find  
 $\zeta_3 \cong v_3/k_BT\kappa_m \gg 1$ in nearly incompressible fluids 
(see Fig.3(d)). To understand the physical meaning 
of $v_i$, let us   prepare  a reference 
 solvent  with $n_1=n_{r1}$ and $n_2=n_{r2}$.  
We  then add a solute  at 
 a small density $n_3$. If   $T$ and $p$ are held fixed,  
 Eq.(A4) in Appendix A gives 
\be 
v_1(n_1-n_{r1})+ v_2(n_2-n_{r2}) + v_3 n_3\cong 0, 
\en 
which holds in  linear order in $n_3$. If $T$, $n_1$,  and $n_2$ 
are held fixed (in a fixed volume), the pressure 
increases by $v_3n_3/\kappa_{\rm m}$ from Eq.(A4).

\subsection{ Kirkwood-Buff theory }

We also introduce the Kirkwood-Buff integrals
\cite{Kirk,Donk,Ben,Lep,Ben-Naim,Shu,Paul,Abbott,Shimizu},
\be  
G_\ab = \int d{\bi r}[g_{ij}(r)-1 ],
\en  
where $g_{i j}(r)$ are the radial distribution functions tending to 1 at large 
$r$. They are  related to the density correlation functions 
$H_\ab(r)= 
\av{\delta{\hat n}_i ({\bi r})
\delta{\hat n}_j ({\bi 0})} $   as 
\be 
H_\ab(r)= 
 n_i n_j [g_\ab (r)-1] + n_i \delta_\ab \delta ({\bi r}). 
\en 
Here, we write   the microscopically defined  number densities  
as  ${\hat n}_i ({\bi r})$ and their deviations 
as  $\delta{\hat n}_i ({\bi r})= {\hat n}_i({\bi r})- n_i$ 
 with caret to avoid confusion with the averages $n_i$.  Then,  
\be 
I_\ab = \int d{\bi r}H_{ij}({\bi r}) 
=  n_i n_j G_\ab + n_i \delta_\ab, 
\en 
which  are the long-wavelength  limits of the structure factors 
$S_{ij}(q) = 
\int d{\bi r}\exp[{{\rm i}{\bi q}\cdot{\bi r}}] H_{ij}({ r})$.
 Hereafter, for any space-dependent variables $\hat{A}({\bi r})$ and 
$\hat{B}({\bi r})$, 
we write\cite{Onukibook}    
\be 
\av{\hat{A}:\hat{B}}\equiv  
\int d{\bi r}[\av{{\hat A}({\bi r})
{\hat B}({\bi 0})}-\av{{\hat A}}\av{{\hat B}}],
\en 
Then,  we have $I_{ij}= \av{\hat n_i: \hat n_j}$.

The  $G_{ij}$ for the solvent species $(i,j=1,2$) 
smoothly tend to those without solute as $n_3\to 0$. 
We  assume that   $G_{i3}$  ($i=1,2, 3$) 
tend to well-defined dilute limits 
  $G_{i3}^0=\lim_{n_3\to 0} G_{i3} $. 
From Eqs.(11) and (A12), we   obtain  
\bea 
\zeta_3&=& - (n_1v_1G_{13}^0+n_2v_2G_{23}^0)/k_BT \kappa_{\rm m},\\
v_3 &=& k_BT \kappa_{\rm m}- (n_1v_1G_{13}^0+n_2v_2G_{23}^0).
\ena 
 
\subsection{Density fluctuations in binary mixtures} 

We here  examine the fluctuations in binary  solvents  
(with  $n_3=0$). Using the deviations 
 $\delta{\hat n}_1$ and  $\delta{\hat n}_2$, we introduce    
   microscopically  fluctuating variables for  the 
volume fraction and the  concentration of the solvent by  
\bea 
\delta{\hat\phi}({\bi r})&=& {v}_1 \delta{\hat n}_1 + 
 {v}_2 \delta{\hat n}_2,\\
 \delta{\hat X}({\bi r})&=& n^{-2}( n_1 \delta {\hat n}_2 -n_2
\delta {\hat n}_1),
\ena 
From  results in Appendix A, we find   
\be 
\av{{\hat\phi}:{\hat\phi}}=  k_BT\kappa_{\rm m},\quad 
\av{{\hat X}:{\hat X}}=  \chi,\quad 
\av{{\hat\phi}:{\hat X}}=0.    
\en 
The  last relation  indicates  {\it orthogonality}  
 between   $\delta\hat\phi$ and $\delta{\hat X}$. 
In terms of $G_{ij}$ for the solvent,  
the compressibility 
$\kappa_{\rm m}$ and   the  concentration 
variance $\chi$ are written as
\bea 
&&\hspace{-1.3cm}
k_BT \kappa_{\rm m}= v_1^2n_1+ v_2^2n_2 +\sum_{i, j=1,2}
 v_in_i v_j n_j G_{ij} \\ 
&&\hspace{-12mm}
 \chi=n_1n_2/n^3  + (n_1^2n_2^2/n^4)  (G_{11}+ G_{22}-2G_{12})\nonumber\\
&=& n^{-1} k_BT (\p X/\p \Delta)_{T,p},   
\ena 
where $G_{ij}$ are those for $n_3=0$ and  $\Delta= \mu_2-\mu_1$.  
The second line of Eq.(23) follows  from  Eq.(A13)\cite{Onukibook}. 
Here,  the  matrix $I_{ij}$ is diagonalized 
by the linear transformations in Eqs.(19) and (20), 
so its determinant is given by   
\be 
I_{11}I_{22}- I_{12}^2= n^4 k_BT \kappa_{\rm m}\chi.
\en 
In nearly incompressible mixtures, 
the fluctuations of $\delta{\hat \phi}$ 
are small, but  those of $\delta{\hat X}$ can  grow 
 due to  molecular clustering\cite{Dixit,Patey1,Dougan,Bag} 
or near  the consolute criticality\cite{Onukibook}.

The scattering intensity  is proportional to 
the structure factor ${I}(q)
=\av{|{\cal A}_{\bi q}|^2}$ 
of a linear combination ${\cal A}= 
Z_1\delta{\hat n}_1+ 
Z_2\delta{\hat n}_2$, where $Z_1$ and $Z_2 $ are constants.  
Here,  ${\cal A}= (Z_1n_1+Z_2n_2)\delta{\hat{\phi}}+ (Z_2v_1-Z_1v_2)
n^2\delta{\hat{X}}$,  so Eq.(21) yields 
the long-wavelength limit\cite{Bha,Donk},  
\be 
{ I}(0)
= (Z_1n_1+Z_2n_2)^2k_BT\kappa_{\rm m}+  (Z_2v_1-Z_1v_2)^2 n^4\chi .
\en 
In water-alcohol mixtures, the  concentration fluctuations 
are enhanced 
on nanometer scales due to the hydrogen-bonding interaction, on which  
a number of  scattering experiments\cite{YKoga,Nishi1,Nishi2,Misawa} 
and  molecular dynamics simulations\cite{Dixit,Patey1,Dougan,Bag} 
were performed. The combination 
 $n(G_{11}+ G_{22}-2G_{12}) $ in  $\chi$  in Eq.(23)  
exhibits   a   maximim  as a function of the alcohol fraction, 
which is   about  $5$ for methanol\cite{Donk}, 
$20$  for ethanol\cite{Donk,Ben,Nishi2,Lep},  
 and     $100$ for TBA\cite{Nishi1,Lep} and 1-propanol\cite{Lep,Misawa},  
  depending   
on the degree of   hydrophobic association 
of alcohol molecules\cite{Bag}.

For  our  model mixture to be 
explained in Sec.III, we  display the mixture quantities 
 in   Fig.1 and 
the normalized solvation chemical potential 
$\nu_3$  in Eq.(2) in Fig.2 
at  $T=300$ K and $p=1$ atm. 
The behaviors  in Fig.1 resemble   those observed  in   water-alcohol 
mixtures, though we do not account for the hydrogen bonding. 

\begin{figure}[t]
\begin{center}
\includegraphics[width=242pt]{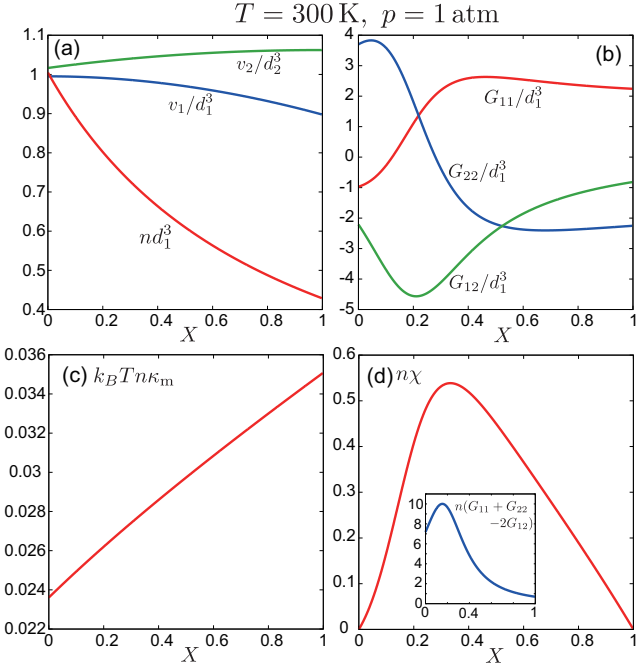}
\caption{ Mixture behaviors 
  without solute  at $T=300$ K and $p=1$ atm  from  MCSL model, 
where $X=n_2/n$ is the cosolvent 
molar fraction with   $n=n_1+ n_2$. Here, our  mixture is in one-phase states 
due to   a strong  attractive interaction between the two species. 
The  hard-sphere  diameters 
are $d_1= 3~{\rm \AA}$ and $d_2=1.3d_1$. 
Thus, $d_1^{3}= 0.0180$ L$/$mol and  $d_1^{-3}= 55.6$ mol$/$L. 
Plotted  are (a) partial volumes $v_1$ and $v_2$ 
divided by $d_1^3$ and 
density $n$ multiplied by $d_1^3$, 
(b) Kirkwood-Buff integrals $G_{ij}$ 
divided by $d_1^3$, (c) normalized compressibility 
$nk_BT \kappa_{\rm m}(\ll 1)$, and 
(d) concentration variance $\chi$ in Eq.(23) multiplied by $n$. 
In  (d),   $n( G_{11}+ G_{22}-2G_{12})$ 
is also shown (inset), whose  maximum 
is about 10 at $X\sim 0.2$. See Sec.III for details of our model. 
}
\end{center}
\end{figure}

\begin{figure}[t]
\begin{center}
\includegraphics[width=242pt]{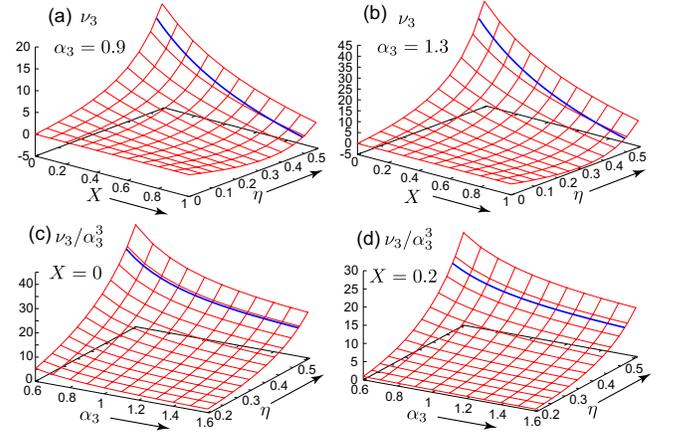}
\caption{Results of the normalized solvation 
chemical potential $\nu_3$  in Eq.(2)  for a hydrophobic solute   
with $\epsilon_{13}=0$ and $\epsilon_{23}/k_B =300$ K at $T=300$ 
K from MCSL model. The solvation chemical 
potential is given by  $ k_BT\nu_3=2.48\nu_3$ kJ$/$mol. 
The diameter ratio  $\alpha_3=d_3/d_1$ 
is   (a) $0.9$ and (b) 1.3  in  
the $X$-$\eta$ plane,   where  
 $\eta=(\pi/6)(n_1 d_1^3+n_2d_2^3)$ is the  hard-sphere  volume fraction. 
Ratio  $\nu_3/\alpha_3^3$ 
is displayed in the $\alpha_3$-$\eta$ plane,  where    $X=n_2/(n_1+n_2)$ 
is   (c) $0$ and (d) 0.2.   For $\alpha_3 >1$, $\nu_3$  
behaves as $\alpha_3^3\propto v_3$.  
The solvent    pressure 
$p_{\rm m}$ in Eq.(6) is 1 atm on blue curves.  See Sec.III. 
    }
\end{center}
\end{figure}

{\subsection{Density fluctuations in ternary mixtures  }}
 
In this subsection, we  superimpose small  density  changes   
$\delta n_i({\bi r})$ on the homogeneous 
averages $n_i$. These $\delta { n}_i$  are coarse-grained variables 
slowly  varying  in space  compared to the potential ranges. 
From Eq.(2)   the second-order change in   $f$  is expressed  as   
\be
{\delta f_{\rm in}} = \frac{1}{2}k_BT \sum_{i.j=1,2,3} 
I^{ij} \delta n_i\delta n_j , 
\en   
where $k_BTI^{ij}=\p^2 f/\p n_i \p n_j =\p \mu_i/\p n_j=\p \mu_j/\p n_i$. 
The  matrix $\{ I^{ij}\}$ is equal to the 
inverse of the variance matrix   $\{ I_{ij}\}$ 
in Eq.(15).
For small $n_3$, Eq.(2) yields 
\be 
I^{i3}=  \nu_{3i} \quad (i=1,2),  \quad 
I^{33}= 1/{n_3}+ U_{33}/k_BT.
\en 
Here,  we neglect the  $n_3$-dependence of 
$I^{ij}$  and $I^{i3}$ ($i, j=1,2$) retaining the first 
diverging term ($\propto n_3^{-1}$) in $I^{33}$, which much simplifies 
the following calculations.

We introduce 
 coarse-grained  deviations of the volume fraction and the concentration 
of the solvent  by 
\be  
\delta{\phi}= {v}_1 \delta{ n}_1 + 
 {v}_2 \delta{ n}_2, \quad  
 \delta{ X}=n^{-2} ({n_1} \delta{ n}_2 -{n_2} \delta{ n}_1), 
\en
which are of the same forms as $\delta{\hat\phi}$ 
and $\delta{\hat X}$ in  Eqs.(19) and (20).   In terms of $\delta\phi$ and 
$\delta X$ the solute-solvent coupling terms in $\delta f_{\rm in}/k_BT$ 
 are  rewritten as 
\be 
( I^{13}\delta { n}_1+I^{23}\delta { n}_2) \delta { n}_3 
=  (\zeta_3  
\delta{\phi}+ g_3 \delta{X}) \delta { n}_3 , 
\en 
where $\zeta_3$ is given in Eqs.(8) and (17). We define 
 the solute-concentration coupling constant $g_3$  by\cite{Ben-Naim,Mochi}    
\be
g_3 = n^2({{ v}_1}{ \nu_{32}}-{ v}_2  { \nu_{31}})=
({\p \nu_3}/{\p X})_{T,p} .
\en 
At fixed $T$ and $p$, Eqs.(A4) and (28) 
give   $v_1 dn_1+ v_2dn_2=0$ and 
 $n^2dX=  v_1^{-1} dn_2= -v_2^{-1} dn_1$, leading to 
  Eq.(30).  In terms of the Kirkwood-Buff integrals 
and  the fluctuation variances, $g_3$   can also be expressed as 
\be  
g_3= -\frac{n_1n_2}{n^2\chi} (G_{23}^0 - G_{13}^0) = 
-\lim_{n_3 \to 0} \frac{\av{{\hat n}_3: 
{\hat X}}}{n_3\av{{\hat X}:{\hat X}}}, 
\en 
with the aid of Eqs.(21) and (A12). 
The difference 
$G_{23}^0-G_{13}^0= k_BT n_2^{-1}( \p\nu_3/\p \mu_2)_{T,p}$ 
represents  the preferential   adsorption 
of a cosolvent  around a solute particle\cite{Abbott}. 
Thus, $g_3=({\p \nu_3}/{\p X})_{T,p}  <0$ 
for  hydrotropic cosolvents.   



For nonvanishing $\zeta_3$ and $ g_3$, 
  ${\delta f_{\rm in}}/{k_BT} $ is written  as 
\be
\frac{\delta f_{\rm in}}{k_BT} =
 \frac{(\delta \phi_{\rm tot})^2}{2k_BT \kappa_{\rm m}} + 
 \frac{ {(\delta X+ \chi g_3\delta n_3 )^2}}{2 \chi}
+ \frac{(\delta n_3 )^2}{2I_{33}}. 
\en 
In the first term,  we  define  the deviation of the  volume fraction 
including a small  solute contribution  as 
\be 
\delta \phi_{\rm tot}= {v}_1 \delta{ n}_1 + 
 {v}_2 \delta{ n}_2 + v_3^{\rm in}\delta n_3,
\en 
where we introduce a solute volume $ v_3^{\rm in}$ by  
\be 
 v_3^{\rm in}= k_BT \kappa_{\rm m}\zeta_3= v_3- k_BT \kappa_{\rm m}.
\en 
From Eq.(11) $v_3^{\rm in}$ 
 is only slightly smaller than $v_3$ for small $\kappa_{\rm m}$.
The second term in Eq.(32) 
indicates  that the solvent composition tends to 
change   by $-\chi g \delta n_3$ with the doping. 
In the third  term,   $I_{33}$ is 
the solute variance  in Eq.(15). From Eq.(27) its 
inverse  is expressed as 
\be 
1/I_{33} = 1/n_3+ U_{33}^{\rm eff}/k_BT.
\en 
The      $U_{33}^{\rm eff}$ is the effective interaction parameter written as  
\bea 
U_{33}^{\rm eff}&=& U_{33}- k_BT \sum_{i,j=
1,2} I_{ij} \nu_{3i}\nu_{3j}\nonumber\\ 
&=& U_{33}  - (v_3^{\rm in})^2/\kappa_{\rm m} 
- k_BT \chi g_3^2 ,
\ena 
where the second line follows from  Eq.(32). 
The second term in  the second line 
is also written as  $-(k_BT)^2 \kappa_{\rm m}\zeta_3^2 $
in terms of $\zeta_3$. The   last two terms are negative 
  representing    the 
solvent-mediated attractive interaction.  Here, 
 $I_{33}^{-1}= ( \p \mu_3/\p n_3)_{T,\mu_1,\mu_2}$ 
from Eq.(A7), so $U_{33}^{\rm eff}$ is  defined as   
\be 
U_{33}^{\rm eff}= \lim_{n_3\to 0} [
(\p \mu_3/\p n_3)_{T, \mu_1,\mu_2}- 
k_BT/n_3].
\en 
which should be compared   with   $U_{33}$ in  Eq.(3).

In  the second line of Eq.(36), the 
 second term  much exceeds $n^{-1}k_BT$ 
 in magnitude for not small solutes  in nearly 
incompressible solvents. 
This contribution is already known  for one-component solvents 
\cite{Pratt1,Widom2,Ashba,Under,Koga1,Liu,Koga2}.
In Sec.III, we shall see  
that the third term can even be larger than 
the second term. Thus,  the right hand side of Eq.(35) 
vanishes when   $n_3$ is equal to a   spinodal  density given by   
 \be 
n_{3}^{\rm spi}=  -k_BT/U_{33}^{\rm eff}.
\en 
We  then have 
$I_{33}= n_3/(1-n_3/n_{3}^{\rm spi})$ and  
\be
G_{33}= 1/(n_{3}^{\rm spi}-n_3), 
\en 
at small nonvanishing $n_3$.  Its   dilute limit is given by 
\be 
G_{33}^0= 1/n_{3}^{\rm spi}= -U_{33}^{\rm eff}/k_BT.
\en 
Here,  $G_{33}>0$   for   $n_3<n_3^{\rm spi}$, 
while  $G_{33}<0$ for  $n_3^{\rm spi}<0$. If  $G_{33}>0$, 
 the interaction among the solute particles is  attractive on the average, 
which can occur  even if the direct interaction  
is repulsive  $(U_{33}>0)$. 
With further increasing $n_3$, we eventually encounter  the unstable regime 
   $n_3>n_3^{\rm spi}>0$,  leading to  
 spinodal decomposition of gas-liquid or 
liquid-liquid phase transition\cite{Onukibook}.    
Similarly,   Roij and Mulders\cite{Roij} calculated 
 demixing spinodal for binary hard-rod mixtures 
using  the second virial expansion  
of the Helmholtz free energy density  as in Eq.(26).

We also notice  that 
the density fluctuations are enhanced 
as $n_3\to n_3^{\rm spi}$ 
in one-phase states. From Eq.(32), 
the concentration variance $\chi_R= \av{{\hat X}: {\hat X}}$  increases  as 
\be 
\chi_R= \chi+ g_3^2\chi^2  n_3/(1-n_3/n_{3}^{\rm spi}), 
\en 
where no  phase-separated droplets are assumed 
and $\chi$ is defined in Eqs.(21) and (23).  
The   second term  can be significant compared to the background $\chi$ for 
$g_3^2\chi  n \gg 1$ even for small $n_3$ (see Fig.3(e) and Eq.(65)).

\subsection{Space-filling condition}

Because  $ v_3^{\rm in}\cong v_3$ 
for small $\kappa_{\rm m}\ll (nk_BT)^{-1}$, the combination 
 $\delta \phi_{\rm tot}$ in Eq.(33) represents   the 
  deviation  of   the total volume fraction  from 1. 
The   space  integral of the  first term in Eq.(32) yields 
the {\it steric free energy}, 
\be 
F_{\rm steric}=   \int d{\bi r}\frac{1}{2\kappa_{\rm m} } 
\big[\sum_{j=1,2,3} v_j \delta n_j\big]^2,  
\en  
where $\delta n_i$ can vary slowly in space. 
This  free energy serves to  
realize   the space-filling $\sum_j v_j n_j=1$ even 
 for inhomogeneous  densities in the limit  of small $\kappa_{\rm m}$. 
See Eqs.(12) and (46) where this 
condition holds  for homogeneous density changes. 
Previously, the steric free energy 
 in the same form 
was assumed  for  polymer mixtures\cite{Onukibook}.

If the composition-dependence 
of $v_i$ is weak  at given $T$ and $p$, 
 the space-filling  holds at any compositions  
with common $v_i$ (depending on $T$ and $p$). 
In our case, this is roughly the case  
in Fig.1(a) and Fig.3(b). 
It is worth noting  that     the Flory-Huggins theory for polymer
mixtures and the Bragg-Williams theory 
of binary alloys\cite{Onukibook,Flory} are based on 
the space-filling {\it assumption}, where 
the volumes of constituent particles (monomers for polymers) 
are   equal to  the cell volume   of  an 
incompressible  lattice.

{\subsection{ Osmotic pressure}} 

For mixture solvents, the  osmotic pressure 
 $\Pi$ is the pressure difference 
$\Pi = p(n_1, n_2,n_3)- p(n_{r1}, n_{r2}, 0)$ 
at fixed chemical potentials 
$\mu_i(n_1, n_2, n_3 )=\mu_i(n_{r1}, n_{r2},0)$ ($i=1,2$),  
where $n_{r1}$ and $n_{r2}$ 
are  the  solvent densities in a reference state with $n_3=0$. 
In the virial expansion 
$
\Pi=k_BT(n_3+ B_2 n_3^2 +\cdots),
$   
 McMillan and Mayer\cite{Mayer} found   
the  relation  $B_2= -G_{33}^0/2$ for one-component solvents. 
Large negative  $B_2$ eventually leads  to solute aggregation.

At fixed $T$, $\mu_1$, and $\mu_2$, we have 
 $d\Pi = n_3 d\mu_3= k_BT n_3 I_{33}^{-1} dn_3 $ 
from  Eq.(A8). Then, 
\bea
&&\hspace{-1cm}\bigg(\frac{\p \Pi}{\p n_3}\bigg)_{T,\mu_1,\mu_2} 
={k_BT}/({1+ n_3 G_{33}})\nonumber\\
&& \hspace{1.3cm}= k_BT (1- n_3/n_3^{\rm spi})      
\ena  
where use is made of Eq.(39) in the second line. Thus,      
\be 
B_2= -1/2n_3^{\rm spi}= -G_{33}^0/2 =U_{33}^{\rm eff}/2k_BT.
\en  
From Eq.(A8) we also find     
\be 
\bigg (\frac{\p{ n_i}}{\p{n_3}}\bigg)_{T,\mu_1,\mu_2}
= \frac{n_i G_{i3}}{1+ n_3 G_{33}} \quad (i=1,2).
\en   
This is integrated to give 
$
n_i-n_{ri}\cong n_{i}G_{i3}^0n_3, 
$  
for small $n_3$ at fixed $T$, $\mu_1$, and $\mu_2$. 
 Therefore, from Eqs.(18) and (34), we find another form of the  
 space-filling condition, 
\be 
v_1(n_1-n_{r1})+ v_2(n_2-n_{r2}) + v_3^{\rm in} n_3\cong 0, 
\en 
where $v_3$ in Eq.(12) is replaced by $v_3^{\rm in}$.

For  nonionic solutes  in  water, 
$B_2$  has been examined    in experiments\cite{Liu} and 
simulations\cite{Pratt1,Widom2,Ashba,Under,Koga1,Koga2}. 
For charged  colloidal particles in a mixture solvent, 
$B_2$  changed from positive to   negative  
on approaching  the  consolute critical point 
before   their near-critical aggregation\cite{Maher,Petit}.

{\subsection{Ostwald coefficient,  Gibbs transfer free energy $\Delta G$,  
and Henry's constant $k_{\rm H}$}}

The solvation chemical 
potential  $k_BT \nu_3$ is measurable  
in two-phase coexistence.
Let phase $\alpha$ be a  a solute-poor  liquid  
and phase $\gamma$ be a solute-rich gas or liquid. 
 The  densities are $n_i^\alpha$  
in phase $\alpha$  and  $n_i^\gamma$  
in phase $\gamma$  ($i=1,2,3)$. 
The  Ostwald coefficient  
  is defined by $L= n_3^\alpha /n_3^\gamma$ for the solute in equilibrium. 
Its dilute limit is written as \cite{Bat} 
\be 
L_0 = \lim_{n_3\to 0}L = \exp[-(\Delta G)_3/ k_BT] ,
\en 
where $(\Delta G)_3$ is  the Gibbs transfer free energy 
(from $\gamma$ to $\alpha$ phase) for a  solute particle. 
Since  $\mu_3$ in Eq.(5)   assumes the same value in the two phases, 
 we find 
\be 
(\Delta G)_3 / k_BT =  \nu_{3}(n_1^\alpha, n_2^\alpha)-
\nu_{3}(n_1^\gamma, n_2^\gamma). 
\en 
If the $\gamma$ phase is a  gas far below the criticality, we obtain 
$(\Delta G)_3  / k_BT \cong \nu_{3}(n_1^\alpha, n_2^\alpha)$, 
since   $\nu_{3}$ is small  in gas.
 For example,   $\Delta G/ k_BT \cong 3.4$  for 
 O$_2$ in  water at $T=300$ K.

  Henry's    constant    
 is   defined in  gas-liquid coexistence. 
Its usual definition is  $k_{\rm H} = f_3/x_3^\alpha$, 
where 
\be 
f_3= k_BT\lambda_3^{-3}
\exp(\mu_3/k_BT) \cong k_BT n_3 e^{\nu_3} 
\en 
is the solute fugacity   and 
 $x_3^\alpha = n_3^\alpha/\sum_j n_j^\alpha$ 
is the solute molar fraction in liquid\cite{Tucker}. 
As $n_3\to 0$ we have 
\be 
k_{\rm H}^0  = \lim_{n_3\to 0}k_{\rm H}=  k_BT n_\ell  
 \exp[{\nu_3(n_1^\alpha, n_2^\alpha)}],
\en 
where $n_\ell= n_1^\alpha + n_2^\alpha$ 
is the liquid density. Thus,   $k_{\rm H}^0 \cong   k_BT  n_\ell/L_0 $ 
far from the   criticality.   

Tucker and Christian\cite{Tucker} furthermore 
obtained  the correction     $k_{\rm H}/ k_{\rm H}^0- 1= -87x_3^\alpha$   
at small   $x_3^\alpha$ for  benzene in  water at 308 K, 
which is weakly hydrophilic with $L_0=3.6$. 
In  Appendix B, $k_{\rm H}/k_{\rm H}^0-1$  
will be  calculated  to linear order in 
$n_3^\alpha$ or $n_3^\gamma$ for one-component solvents as   
\bea 
&&\hspace{-10mm}{k_{\rm H}}/{k_{\rm H}^0} -1 
=( 2B_2^\alpha+ 1/n_1^\alpha -2v_3^\alpha) n_3^\alpha + 
k_BT\kappa_\alpha n_3^\alpha\nonumber\\
&&\hspace{1cm} + v_3^\alpha ({ 
n_1^\alpha  - n_1^\gamma} )^{-1} (
{ n_1^\alpha n_3^\gamma -n_1^\gamma n_3^\alpha}),
\ena 
where  $\kappa_\alpha$,  $v_3^\alpha$, and $B_2^\alpha$ 
are the liquid values of  $\kappa$,  $v_3$, and $B_2$. 
For benzene, we have  $n_\ell B_{2}^\alpha \sim -50$  
 and $n_\ell  v_3^\alpha \sim 10$, so     
 the  first term  is dominant in 
Eq.(51) and   can well  explain   
 the observed  correction\cite{Tucker}, 
as was shown by 
Rossky and  Friedman\cite{Rossky}.   However, 
for hydrophobic solutes with $L_0\ll 1$, 
we have    $ k_{\rm H}/k_{\rm H}^0 -1 \cong 
 v_3^\alpha n_3^\gamma$. 
See Appendix B for the correction $L/L_0-1$  
of the Ostwald coefficient.

\section{Numerical results on solvation }

In the density functional theory\cite{Evansreview,Ox,Ox1},   
  use   has been made of the Carnahan-Starling (CS) model  
for pure fluids\cite{Car0} and 
the  Mansoori-Carnahan-Starling-Leland  (MCSL) 
  model   for  mixtures\cite{Man} (see  Appendices C and D).
These models  provide 
the equation of state for hard spheres 
 in agreement with simulations\cite{Man}.  
In our previous paper\cite{OkaD}, 
we used  the MCSL model for water-oxygen 
mixtures. In this paper, 
 we use it  for  ternary mixtures.

\subsection{MCSL  model and van der Waals model}

For the  coarse-grained  smooth  densities $n_i$, 
we write  the  Helmholtz free energy  density as  
\be 
f= \sum_{i=1,2,3} k_B T n_i
 [\ln (n_i\lambda_i^3)-1]+ f_h + f_a,   
\en  
where  $f_h$ is the hard-sphere  part  
and  $f_a$ is the attractive part. 
The  potential 
from the boundary walls\cite{Evansreview} is not written. 
 Each  species    has    a hard-sphere  diameter $d_i$. 
We write 
the diameter ratios  as 
\be 
\alpha_2=d_2/d_1, \quad \alpha_3= d_3/d_1.
\en   
In the literature\cite{Evansreview,Ox,Ox1}, $f_a$ 
   has   the pairwise    form   
with   Lennard-Jones potentials 
$\phi_{ij}(r)$ for $r>d_{ij}= (d_i+ d_j)/2$. 
In this paper, 
 we use the simple   van der Waals form\cite{Onukibook},  
\be
 f_a =  -\frac{1}{2} \sum_{i,j=1,2,3} w_{ij} n_i n_j.   
\en 
The coefficients   $w_{ij}$ are related to the hard-sphere diameters 
and the Lennard-Jones energies $\epsilon_{ij}$  of $\phi_{ij}$ by\cite{Ox,Ox1}   \be  
w_\ab =-\int_{r>d_{ij}} \hspace{-1mm}
d{\bi r}\phi_\ab(r)= \frac{4}{9} \sqrt{2}\pi\epsilon_\ab ( d_{{i}}+d_j)^3 .
\en  
In our scheme, the chemical potentials  are written as 
\be 
\mu_i= k_BT \ln (n_i\lambda_i^3) +\mu_{hi}- \sum_j w_{ij} n_j ,
\en   
See    Eq.(D3) for  the hard-sphere  part   $\mu_{hi}=\p f_h/\p n_i$.

As   $n_3\to 0$, we calculate $\nu_3 $ 
and $U_{33} $ in Eq.(2) as  
\bea 
&&\hspace{-1cm}
\nu_3 (n_1,n_2)=  [\lim_{n_3\to 0} \mu_{h3} -w_{13}n_1- w_{23}n_2]/k_BT,\\
&& \hspace{-11mm} 
U_{33}(n_1,n_2)=  \lim_{n_3\to 0} (\p \mu_{h3}/\p n_3) - w_{33}.
\ena 
Here,  $\nu_3$ and $U_{33}$ are complicated 
functions of  $\eta= \pi(n_1d_1^3+ n_2d_2^3)/6$, 
$X$, $\alpha_2$, and $\alpha_3$ (see Appendix D). 
They increase steeply with increasing $\eta \gs 0.5$ 
and behave as $\alpha_3^3\propto v_3$ and $\alpha_3^6\propto v_3^2$, 
respectively,  for $\alpha_3 >1$. 
In particular, for    one-component 
solvents ($n_2=0$) we  obtain\cite{Lada}   
\bea 
&&\hspace{-4mm}
 \nu_3(n_1,0)= (3\alpha_3 +6\alpha_3^2-\alpha_3^3 )u_1 + 
3\alpha_3^2u_1^2 +\alpha_3^3 u_1^2(4+2u_1) 
 \nonumber\\
&&\hspace{-2mm} 
- (\alpha_3-1)^2 (2\alpha_3+1) \ln (1- \eta_1)-w_{13}n_1/k_BT,
\ena 
where $\eta_1 = (\pi d_1^3/6) n_1$ and   
$u_1=\eta_1/(1-\eta_1)$.

{\subsection{ Selected parameter values for mixture solvent}}

\begin{figure}[tbp]
\begin{center}
\includegraphics[width=242pt]{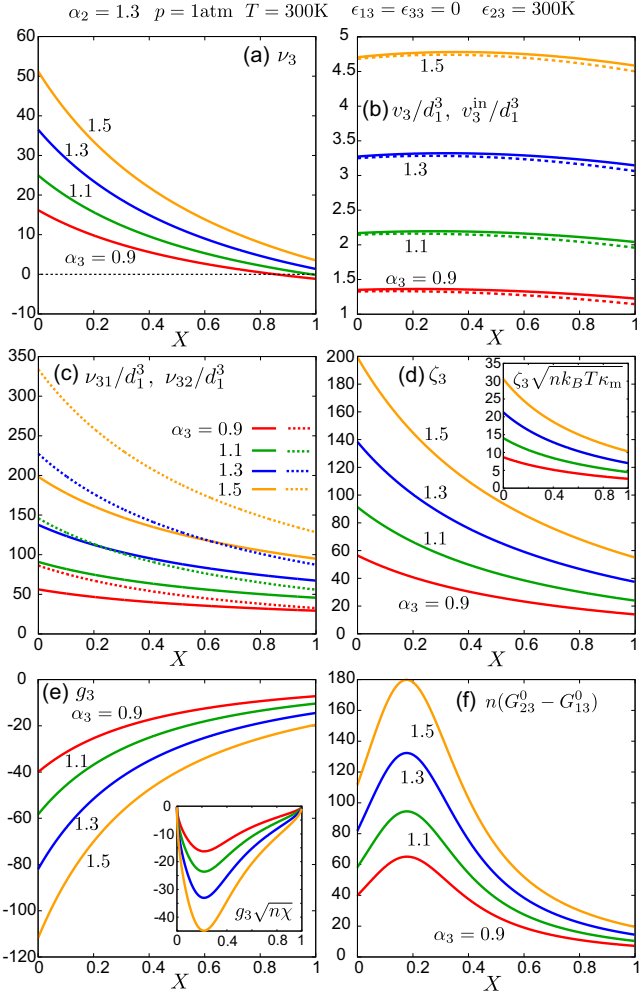}
\caption{ Solvation quantities  
vs  the cosolvent molar fraction  $X$ 
for a  solute  with  diameter ratio 
$\alpha_3=d_3/d_1= 0.9, 1.1, 1.3$, and 1.5  
at $T=300$ K and $p=1$ atm, where 
the solvent  is characterized by Eqs.(60) and (62) with 
 $\epsilon_{13}=0$ and $\epsilon_{23}/k_B=300$ K.   
  Plotted are (a) $\nu_3$,   (b) solute partial volume 
 $v_3$ (bold lines) and another volume 
$v_3^{\rm in}$ in Eq.(34) (dotted lines)  
divided by $d_1^{3}$, which are very close, 
(c) density-derivatives $\nu_{31}$ (bold lines) 
 and $\nu_{32}$ (dotted lines) in Eq.(7) 
divided by  $d_1^{3}$,  (d) solute-density coupling 
coefficient  $\zeta_3$ in Eq.(8),
(e) solute-concentration coupling 
coefficient $g_3$ in Eq.(30), 
and (f) degree of preferential 
adsorption  $n(G_{23}^0-G_{13}^0)$ appearing in Eq.(31). 
In  insets in (d) and (e),  
$\chi_3 (nk_BT \kappa_{\rm m})^{1/2}$ 
and $g_3 (n\chi)^{1/2}$  are 
typical sizes of $\zeta_3\delta\phi$ 
and $g_3\delta X$  in Eq.(29). 
   }
\end{center}
\end{figure}

Supposing   ambient  water close to  its gas-liquid  coexistence 
as  a reference state  of  the first species, we  set   
\be 
d_1= 3~{\rm \AA}, \quad 
\epsilon_{11}/k_B= 588.76~ {\rm K}, 
\en 
where  $w_{11}/\epsilon_{11} d_1^{3} = 15.80$  from Eq.(55).   
We choose  $\epsilon_{11}$  such that 
the coexistence (saturated vapor)  pressure   at $T=300$ K    
is equal to its experimental one   $ p_{\rm cx}^0=
0.031$ atm (see Appendix C). 
The density in the reference state   $n_r$ 
and the coexisting  liquid and gas water densities, 
$n_\ell$ and $n_g$, 
at $T=300$ K are calculated as  
\bea 
&&
 n_r=1.0049d_1^{-3},\quad 
n_\ell=n_r -1.5 \times 10^{-5}n_r, \nonumber\\
&& n_g =p_{\rm cx}^0/k_BT=  2.1 \times 10^{-5}n_r. 
\ena 
The hard-sphere  volume fraction  
is $\eta_1 =  \pi n_r d_1^3/6 =  0.526$ for this  $n_r$.  
The  $n_\ell$ here is larger than the corresponding  density 
 of  real water  by $10~\%$. 
The  compressibility   is 
 $1.71 \times 10^{-4}/$MPa $= 0.023 /n_rk_BT$ in  the reference state, 
while the experimental one is $4.5 \times  10^{-4}/$MPa.

We choose the parameters of the second species 
such that   the  mixture solvent  remains 
in one-phase states at any $X$ in ambient conditions. 
We thus set     
\be
d_2 = 3.9~{\rm \AA}, \quad \epsilon_{12}/k_B=550~{\rm K}  ,
\quad \epsilon_{22}/k_B=500~{\rm K}, 
\en
for which   $w_{12}/\epsilon_{11}d_1^3=22.44$ and 
 $w_{22}/\epsilon_{11}d_1^3=29.47$. 
Here, $\epsilon_{12}$ and $\epsilon_{22}$ are 
  close to $\epsilon_{11}$.  In Fig.1, we have  
displayed the   mixture properties.  
We further make remarks. (i) As 
$n_2 \to 0$,  the second species has  the Gibbs transfer free energy 
 $(\Delta G)_2 \cong -7.8k_BT$  at $T=300$ K, so it is hydrophilic. 
(ii)  At $T=300$ K, the gas-liquid coexistence  pressure 
$p_{\rm cx}^0$ without solute 
is calculated as $ 0.031$, $0.075$, $0.089$, $0.10$, 
$ 0.11$, and $ 0.13$ in units of  atm, where  
the molar fraction $X^\alpha$ of the second species in liquid is  
 0, 0.2, 0.4, 0.6,0.8, and 1.0, respectively. 
These low pressures stem from the strong 
hydrophilicity of the second spiecies 
(see Eq.(B3) and sentences below it).        
Note that the coexistence  
pressure of  real water-alcohol mixtures  
is very low at room temperatures.

We then add a small amount of a 
hydrophobic solute. In most cases to follow, 
we set  $\epsilon_{13}=\epsilon_{33}=0$ 
and $\epsilon_{23}= 300$ K,
for which the solute  interacts with   
the two solvent species  differently, 
resulting in    a large  solute-concentration coupling. In Fig.2, 
we have  shown the overall behaviors of $\nu_3$ 
in the $X$-$\eta$ plane at fixed $\alpha_3$  and 
 $\nu_3/\alpha_3^3$ 
in the $\alpha_3$-$\eta$ plane at fixed $X$.
The $\nu_3$ increases 
steeply with increasing $\eta$ 
for $\eta\gs 0.5$,  decreases with increasing $X$,   
and behaves as  $\alpha_3^3$ for $\alpha_3 \gs 1$ (see  
Appendix D for analytic results). The asymptotic 
  relation $\nu_3 \sim \alpha_3^3$ is natural. 
The same result follows  from 
 the simple van der Waals  model of fluid mixtures\cite{Onukibook}, 
where    the steric free energy density 
is given by $-k_BT (\sum_i n_i)  \ln (1-\sum_i v_i^0 n_i)$, 
with  $v_i^0$ being hard-sphere volumes.

In  Fig.3, we plot   quantities 
related to $\nu_3$  vs  $X$. 
Displayed are   (a) $\nu_3$,   (b)  $v_3$ in Eq.(11) 
and $v_3^{\rm in}$ in Eq.(34),
(c)  $\nu_{31}$  
 and $\nu_{32}$ in Eq.(7),  (d) $\zeta_3$ in Eq.(8),
(e) $g_3$ in Eq.(30), and (f) 
 $n(G_{23}^0-G_{13}^0)$. 
These quantities increase  as $\alpha_3^3$  
with increasing  $\alpha_3$. In the insets in (d) and (e), 
 $\zeta_3 (nk_BT \kappa_{\rm m})^{1/2}$ 
and $g_3 (n\chi)^{1/2}$ indicate 
large sizes of $\zeta_3\delta\phi$ 
and $g_3 \delta X$ in Eq.(29).   In (d) and (e), the solute-solvent 
coupling coefficients $\zeta_3$ and $g_3$ are both large. 
In (f),  $n(G_{23}^0-G_{13}^0)$ 
is  peaked at $X\sim 0.2$. Previously,  
 Booth {\it et al.}\cite{Abbott} 
found a maximum of    $G_{23}^0 - G_{13}^0$  as a function of the 
hydrotrope density $n_2$ for various  solutes (drugs) 
in water-hydrotrope mixture solvents.

We should have  $|g_3|\gg 1$ and      
$  g_3^2n\chi\gg 1$ for relatively large 
hydrophobic solutes in  water-hydrotrope  solvents, 
especially in the presence of pre-Ouzo 
aggregates as thermal fluctuations\cite{Zemb,Zemb1,Zemb2,Hori,Ani2}.  
For  water(1)-ethanol(2)- octanol(3),  
 results of   molecular dynamics simulation\cite{Zemb2,Hori} 
lead to   rough estimatations,  $n(G_{13}^0-G_{23}^0)= 100-300$, 
$-g_3=  20-60$,  and $g_3^2 n\chi = 300-3000$,  at $X=0.2$.  
For smaller (less hydrophobic) methane at $X\sim 0.2$, 
$g_3$ was numerically  about $-5$ in  water-methanol\cite{Mochi} 
and about $-15$  in water-TBA\cite{Lee}.

{\subsection{ $n_3^{\rm spi}$,  $G_{33}^0$, and $B_2$ 
for mixture solvent}}

\begin{figure}[tbp]
\begin{center}
\includegraphics[width=242pt]{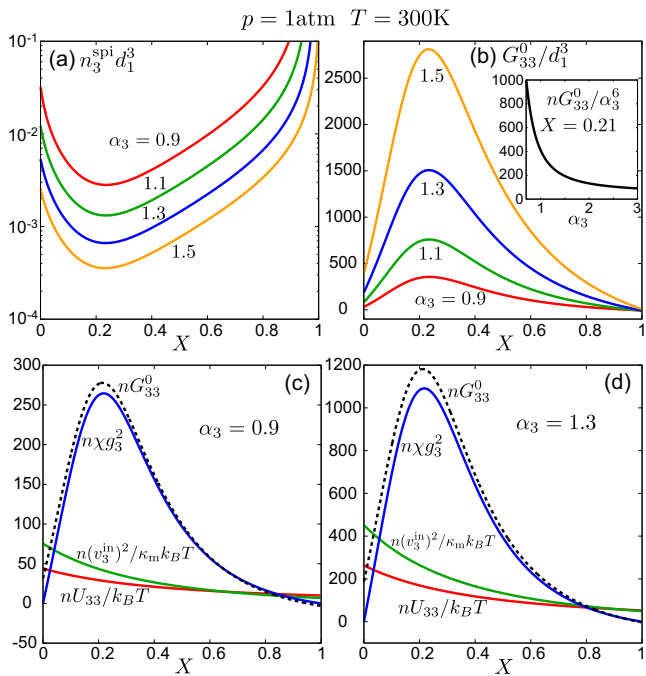}
\caption{ Results as functions of the cosolvent molar fraction  $X$ 
for a solute with $\epsilon_{13}=\epsilon_{33}=0$ 
and $\epsilon_{23}/k_B=300$ K  at $T=300$ K and $p=1$ atm.  
(a) Spinodal solute density 
$n_3^{\rm spi} (= -k_BT/ U_{33}^{\rm eff})$ in Eq.(38) 
 multiplied by $d_1^3$ and  (b)  Kirkwood-Buff integral 
$G_{33}^0(=1/n_3^{\rm spi})$ in the dilute limit in Eq.(40) 
divided by $d_1^3$, where the 
  diameter ratio $\alpha_3=d_3/d_1$ is $ 0.9, 1.1, 1.3$, and 1.5. 
In inset in  (b), $n  G_{33}^0/\alpha_3^6$ is plotted at $X=0.21$. 
In   (c)   and (d), 
three contributions to $n U_{33}^{\rm eff}/k_BT
 (=-nG_{33}^0)$ in Eq.(36) are compared  for 
$\alpha_3= 0.9$  and 1.3, respectively, 
where concentration contribution 
$n\chi g_3^2$ dominates  and is nearly equal to 
$nG_{33}^0)$ for $X\gs 0.05$. 
 }
\end{center}
\end{figure}

The   spinodal solute density $n_3^{\rm spi}$,  
 the Kirkwood-Buff integral $G_{33}^0$,  and  
the second osmotic virial coefficient   $B_2$  
are related as  $G_{33}^0= -2B_2=1/n_3^{\rm spi}$   in Eqs.(40) and (44). 
In Fig.4, we plot (a) $n_3^{\rm spi}$ and (b)  
$G_{33}^0$ vs $X$, where the former (latter) 
decreases (increases) with increasing   $\alpha_3$. 
The minimum of $n_3^{\rm spi}d_1^3$ is small, 
which is 
$2.82\times 10^{-3}$ at $X=0.236$ 
and is  $6.63\times 10^{-4}$ at $X=0.234$ for $\alpha_3=1.3$.   
In (c) and (d), 
we decompose $ U_{33}^{\rm eff}/k_BT (= -G_{33}^0) $ into 
 $U_{33}/k_BT$,  $-(v_3^{\rm in})^2 /k_BT\kappa_{\rm m}$, 
and $-g_3^2\chi$ from  Eq.(36) and compare them 
for $\alpha_3=0.9$ and 1.3. For these examples, 
the concentration part dominates for not very small $X(\gs 0.05$) 
and   the sum of the first two terms   
nearly vanishes for $X\gs 0.8$. Thus,  for $g_3^2 n\chi \gg 1$ 
and  $X\gs 0.05$,  we find   
\be 
n_3^{\rm spi}= 1/ G_{33}^0
\cong (g_3^2\chi)^{-1}.
\en

{\subsection{ $n_3^{\rm spi}$,  $G_{33}^0$, and $B_2$ for one-component  solvent}}

In nearly incompressible,   one-component solvents ($X=0$),  
there is no  concentration 
part   in Eq.(36), but the direct interaction  
 $U_{33} $ and the solvent-mediated one  $-(v_3^{\rm in})^2 /\kappa_{\rm m}$ 
are still  large in magnitude ($\gg k_BT/n_1$) for 
not small solutes. 
They  largely cancel  but their difference can  
 give large positive  $G_{33}^0=-2B_2$. 
  The resultant attractive 
interaction leads  to  the well-known assembly 
of hydrophobic particles 
in ambient water\cite{Amo1,WidomH,Chandler}.
Furthermore, for  $n_3>n_3^{\rm spi}=1/G_{33}^0$,  
 spinodal decomposition of  gas-liquid phase 
transition occurs.  Previously,  these two contributions to $U_{33}^{\rm eff}$ 
(or to $B_2$)  
have  been calculated separately   for one-component 
solvents\cite{Widom2,Koga1}, but  there has been no  analysis of their 
dependences on the solute size and 
the solute-solvent attractive interaction.  

In Fig.5, we plot 
$n_3^{\rm spi}$, $G_{33}^0$, $G_{33}^0/\alpha_3^6$, 
and $(v_3^{\rm in})^2/\kappa_{\rm m}U_{33}$ vs  $\alpha_3$ 
at  $X=0$, where  $\epsilon_{13}/k_B$ is  $0$, 250,  and 400 K. 
In  these cases, $n_3^{\rm spi}$ rapidly decreases to very 
small values with increasing  $\alpha_3$. 
From (c) its behaves as  
\be 
n_3^{\rm spi} =1/G_{33}^0\sim  A_0 n_1  \alpha_3^{-6}\quad (X=0),  
\en  
where $A_0 =10^{-1}-10^{-2}$ for not large 
$\epsilon_{13}(<\epsilon_{11})$. 
See Appendix D for analytic results.   
We can also see that  $G_{33}^0$  decreases  
with increasing $\epsilon_{13}$, as it should be the case. In (c), it is  
 negative in the range $\alpha_3<1.22$ 
for   $\epsilon_{13}/k_B=400$ K, 
while it is positive   for any $\alpha_3$ at 
 $\epsilon_{13}=0$ (hard-sphere solutes)\cite{Pratt1}.
In (d), the ratio of the two contributions to   $U_{33}^{\rm eff}$ 
tends to a constant for each $\epsilon_{13}$. 
\begin{figure}[t]
\begin{center}
\includegraphics[width=235pt]{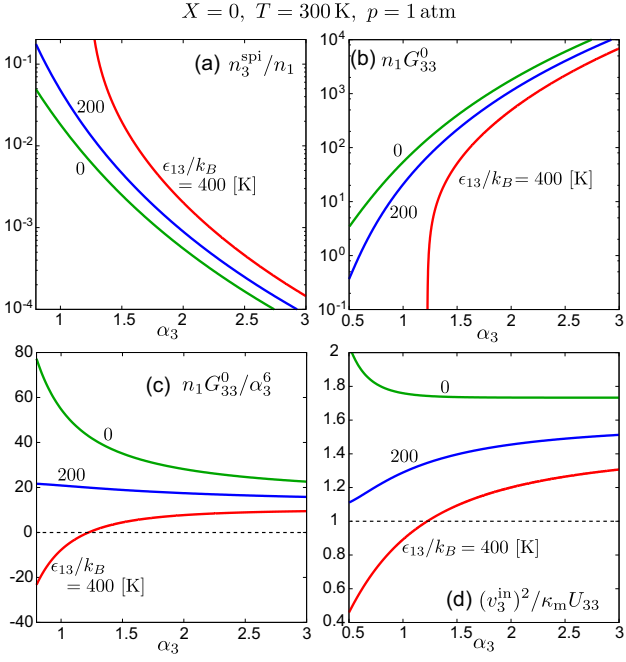}
\caption{ Results  in 
   one-component solvent ($X=0$) slightly outside 
the coexistence curve  at $T=300$ K and $p=1$ atm.   
 As functions of diameter ratio $\alpha_3=d_3/d_1 (\le 3)$,  
displayed are (a) spinodal solute  density 
$n_3^{\rm spi}(= -1/2 B_2)$ divided by $n_1$,  
(b) Kirkwood-Buff integral in the dilute limit 
$G_{33}^0(= -2B_2)$ multiplied by $n_1$, 
(c) $n_1G_{33}^0/\alpha_3^6$, 
and (d)  $(v_3^{\rm in})^2/\kappa_{\rm m}U_{33}= 1- U_{33}^{\rm eff}/
U_{33}$ (see Eq.(36)). 
Here, $\epsilon_{12}/k_B$ takes three values ($0$, 250, and 400 K), 
which is 0 in the other figures in this paper. 
In (c), $G_{33}^0 \propto \alpha_3^6$ for large $\alpha_3(\gs 2)$ 
 and $G_{33}^0 $  changes its sign 
at $\alpha_3=1.22$ for  
 $\epsilon_{12}=400k_B$. In (d), the ratio tends to 
a constant larger than 1 with increasing $\alpha_3$, leading to  $G_{33}^0>0$. 
}
\end{center}
\end{figure}

\begin{figure}[tbp]
\begin{center}
\includegraphics[width=240pt]{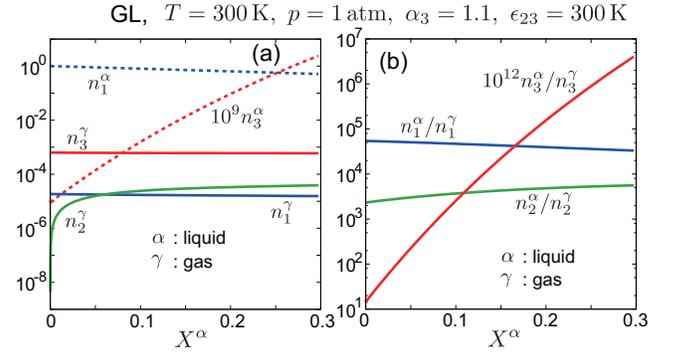}
\caption{ Results of solute-induced 
 gas-liquid coexistence at $T=300$ K and $p=1$ atm,  
where $\alpha_3=1.1$, $\epsilon_{13}=\epsilon_{33}=0$, 
 and $\epsilon_{23}/k_B=300$ K.
(a) Densities in liquid $n_i^\alpha$ (dotted lines) 
and those in gas $n_i^\gamma$ (bold lines) in units of $d_1^{-3}$ 
vs the cosolvent molar fraction in liquid 
$X^\alpha= n_2^\alpha/(n_1^\alpha+ n_2^\alpha) (\le 0.3)$, where 
the solute density in liquid $n_3^\alpha$ is extremely small. 
  (b) Density ratios $n_i^\alpha/n_i^\gamma$ vs $X^\alpha$, where 
those for the solvent ($i=1,2$)  
 depend  on $X^\alpha$  rather weakly but that 
for the solute ($=$the Ostwald coefficient $L$)  
 depends  on $X^\alpha$ very strongly.}
\end{center}
\end{figure} 
\begin{figure}[t]
\begin{center}
\includegraphics[width=242pt]{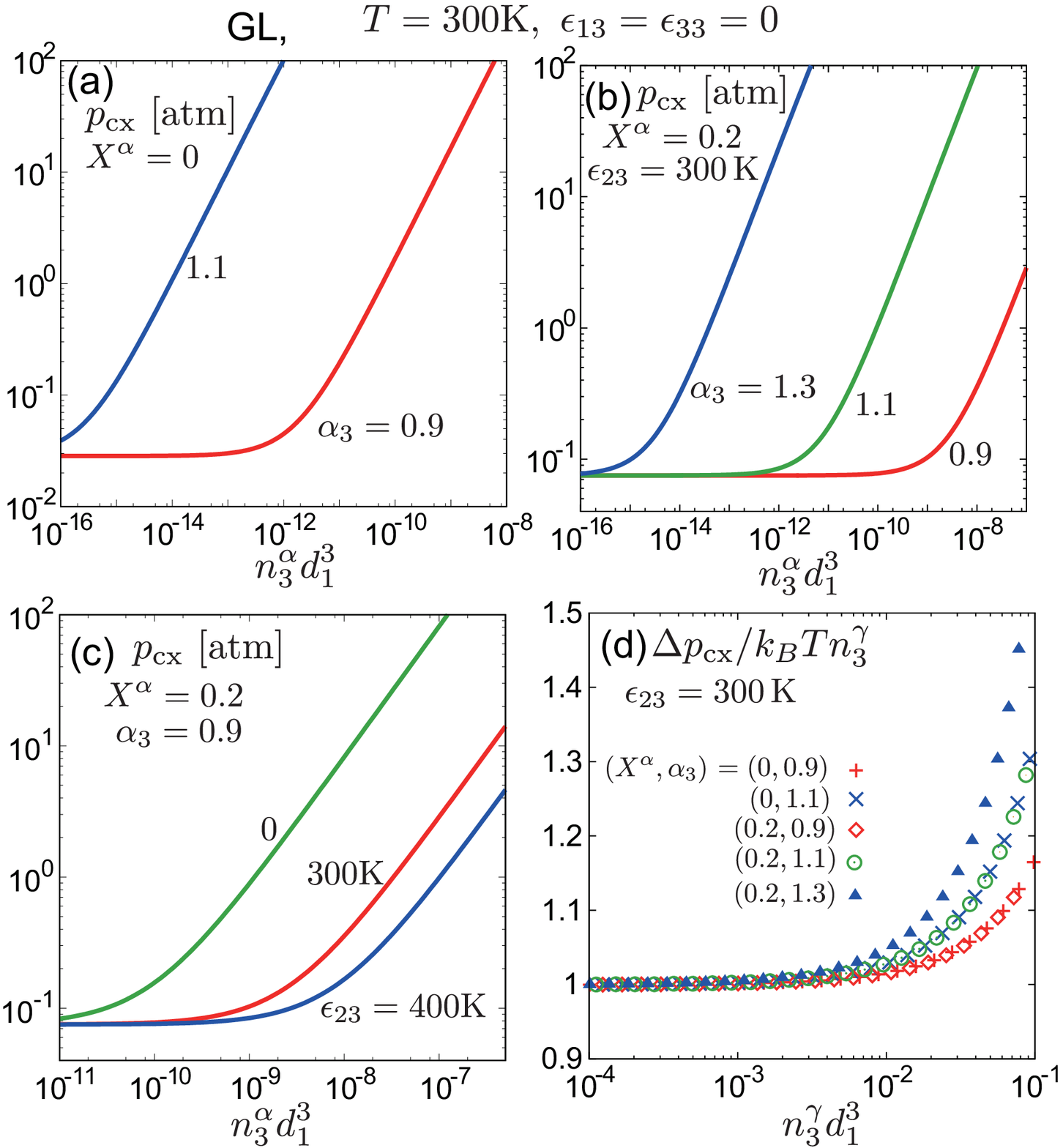}
\caption{ Gas-liquid coexistence 
pressure  $p_{\rm cx}$ in units of atm 
with addition of a hydrophobic solute with 
$\epsilon_{13}= \epsilon_{33}=0$ at $T=300$ K. 
As functions of $n_3^\alpha d_1^3$ on logarithmic scales,  $p_{\rm cx}$  is 
plotted   (a) for   $X=0$ 
with  $\alpha_3=0.9$ and 1.1, 
(b)  for  $X^\alpha=0.2$ with 
$\alpha_{3}= 0.9, 1.1$, and 1.3 at $\epsilon_{23}/k_B=300$ K, 
and (c)  for   $X^\alpha=0.2$ 
with $\epsilon_{23}/k_B 
=0, 300,$ and 400 K  at $\alpha_3=0.9$.  Here, 
$ p_{\rm cx} \to p_{\rm cx}^0 (\ll 1$ atm)   as 
$n_3^\alpha \to 0$
(d) Ratio $\Delta p_{\rm cx}/k_BT n_3^\gamma$ 
 vs $n_3^\gamma d_1^3$ for five cases in 
(a) and (b) on a semi-logarithmic scale, where   
$\Delta p_{\rm cx}= p_{\rm cx}- p_{\rm cx}^0$ 
is the pressure increase due to  solute.  
  }
\end{center}
\end{figure}

\section{Phase separation with a solute} 

We now examine  gas-liquid and liquid-liquid 
coexistence  induced by a hydrophobic solute in our 
mixture  solvent characterized by Eqs.(60) and (62) at $T=300$ K. 
We  require   that  the chemical potentials $\mu_i$ and the pressure 
$p$ assume common values in  coexisting two phases, so we treat macroscopic 
phase separation. The resultant 
densities   are  written  
 as $n_i^\alpha$ in the solute-poor liquid 
phase and   $n_i^\gamma$ in the solute-rich  phase.  
 The surface tension is   included  in Sec.IVE in discussions 
of nucleation.

\subsection{Solute-induced gas-liquid  phase separation}

In our gas-liquid transition, the solute is 
 squeezed out of the liquid, while  
 $n_1^\gamma$ and $ n_2^\gamma$  remain small. 
In Fig.6, we show the densities in   gas-liquid coexistence 
by varying      the  molar fraction 
$X^\alpha= n_2^\alpha/(n_1^\alpha+ n_2^\alpha)$ 
at fixed  pressure    $p=p_{\rm cx}=1$ atm with $\alpha_3=1.1$. 
Gas-liquid coexistence in  ternary mixtures  
is uniquely determined for each  given $T$, $p$, and $X^\alpha$ 
For our parameter choice,   $(\Delta G)_3/k_BT $ in Eq.(48) 
  decreases  with increasing $X^\alpha$, 
which   is $ 25.0 $ at  $X^\alpha=0$ 
and  $12.3 $ at $X^\alpha=0.3$ for $\alpha_3=1.1$  
(16.19 at  $X^\alpha=0$ 
and   9.81  at $X^\alpha=0.2$ for $\alpha_3=0.9$). 
The ratio $n_2^\alpha/n_2^\gamma$ 
  stays close to its dilute limit ($n_2\to 0$) 
given by  $\exp(-(\Delta G)_2/k_BT)\sim e^8$. 
Thus, the second species plays the role of a cosolvent 
improving the solute solubility.

In Fig.7,  we show that 
the pressure   $p_{\rm cx}$ increases 
with increasing $n_3^\alpha= L_0 n_3^\gamma$
 in  gas-liquid coexistence   fixed    $T$ and $X^\alpha$. In (a), 
for the one-component solvent case, 
 we plot  $p_{\rm cx}$ vs $n_3^\alpha$   for $\alpha_3=0.9$ and 1.1. 
Also  for  $X^\alpha=0.2$, 
we plot    $p_{\rm cx}$  vs $n_3^\alpha $  
 for  three values of 
 $\alpha_{3}$ at $\epsilon_{23}/k_B=300$ K in (b) 
and  for  three values of  $\epsilon_{23}$  at $\alpha_3=0.9$ in (c).  
For each  $n_3^\alpha$, 
  $p_{\rm cx}$ largely increases  with 
increasing the solute hydrophobicity. 
In (d), the ratio $ (p_{\rm cx}- p_{\rm cx}^0)/k_BT n_3^\gamma$  
tends to 1  for $n_3^\gamma d_1^3 \ll 1$ (see  Eq.(65)), 
while it  is   between $[1.1, 1.5]$ at   $n_3^\gamma d_1^3\sim 0.1 $ 
 due to the solute-solute repulsion.
   Larger  $n_3^\gamma $ can be obtained  at higher pressures (up to 
  $0.337 d_1^{-3}$ in  Fig.9(b)). Furthermore, 
if we  include the   attractive interaction 
among the solute particles ($\epsilon_{33}>0$),  
$n_3^\gamma$  can be increased 
up to a liquid density.

In the dilute limit $n_3^\gamma \ll  d_1^{-3}$, 
 $p_{\rm cx}$  is simply given by\cite{OkaD,nano}   
\be 
p_{\rm cx}- p_{\rm cx}^0=  k_BT n_3^\gamma
= k_BT L_0^{-1} n_3^\alpha=f_3 \quad ({\rm GL}),
\en 
where  $p_{\rm cx}^0$ is the coexistence pressure 
without solute,  $L_0$ 
is the Ostwald coefficient in Eq.(47), 
and $f_3$ is the solute fugacity in Eq.(49).  
 In accord with Eq.(65),  
  both $L_0$ and  $n_3^\alpha$ increase  
with increasing $X^\alpha$  at fixed  $p_{\rm cx}$ 
in  Fig.6.

We  can now  discuss  metastability of  reference  liquid states  
with respect to the gas-liquid  transition 
by   increasing its  
solute density ${\bar n}_3$  
at given  pressure ${\bar p}~(>p_{\rm cx}^0)$.  
 In this  isobaric condition, 
  the solvent densities decrease 
according to Eq.(12).    From Eq.(65), the   liquid is 
 stable for ${\bar n}_3<n_3^{\rm b}({\bar p})$ but is 
metastable  for ${\bar n}_3>n_3^{\rm b}$, where  
\be  
n_3^{\rm b}=
 L_0( {\bar p}- p_{\rm cx}^0)/k_BT \quad ({\rm GL}).
\en 
Recall that instability  
occurs  for   ${\bar  n}_3> n_3^{\rm spi}$, where 
$n_3^{\rm spi}(>n_3^{\rm b})$ is the  spinodal solute density in Eq.(38).  
For very small  $L_0$ and ${\bar p}-p_{\rm cx}^0$, 
$n_3^{\rm b}$ can be  extremely small. 
We may also   change  $\bar p$ fixing  ${ \bar n}_3$, where 
 the liquid  is  metastable  for ${\bar p}<p_{\rm b}({\bar n}_3)$. 
This    $p_{\rm b}$ is     the 
 bubble-point pressure,  
\be 
p_{\rm b} = p_{\rm cx}^0+  k_BT L_0^{-1} {\bar n}_3 \quad ({\rm GL}),    
\en 
which  is equivalent to   Eq.(65). 
For example,  $p_{\rm b}\sim 
10^3$ atm  if  the solute   molar fraction is 
of order  $ L_0 $. 
In  pure water at $T\sim 300$ K, 
 in contrast,    bubble nucleation 
was observed 
for   large negative pressures ($\sim 
-10^3$ atm)\cite{Angell,Caupin}.

\begin{figure}[t]
\begin{center}
\includegraphics[width=240pt]{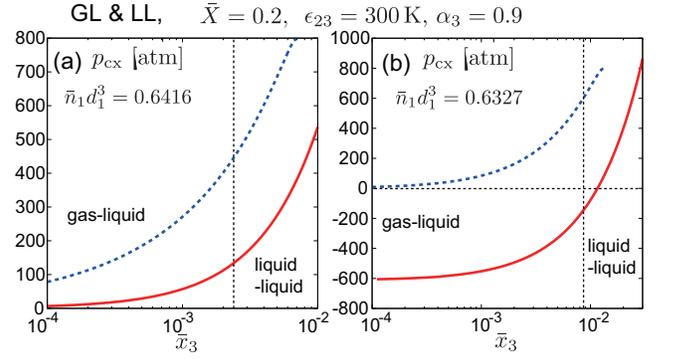}
\caption{ Coexistence pressure $p_{\rm cx}$ vs average solute 
fraction ${\bar x}_3= N_3/(N_1+ N_2)$ 
 at   fixed particle numbers $N_i={\bar n}_i V$ 
 in a  fixed volume $V$, where  
gas-liquid  (blue  dotted lines)  
and liquid-liquid (red bold lines) transitions 
 occur. Here,   ${\bar n}_1 d_1^3$ is 
(a) $0.6416$ and (b) $0.6327$, where 
  ${\bar X}=N_2/(N_1+N_2)=0.2$, 
 $\alpha_3=0.9$,  $\epsilon_{13}=\epsilon_{33}=0$, and $ 
\epsilon_{23}/k_B=300$ K.
On the liquid-liquid branch, $p_{\rm cx}$ approaches 
 (a) $p_{\rm cx}^0=0.075$ atm   
and (b)  $-613.5$ atm as ${\bar x}_3 \to 0$. 
On the gas-liquid branch, $p_{\rm cx}-p_{\rm cx}^0
 \cong k_BT n_3^{\gamma}$ 
for $n_3^{\gamma} d_1^3\ll 1$ as in Eq.(65). 
}
\end{center}
\end{figure}

\begin{figure}[t]
\begin{center}
\includegraphics[width=240pt]{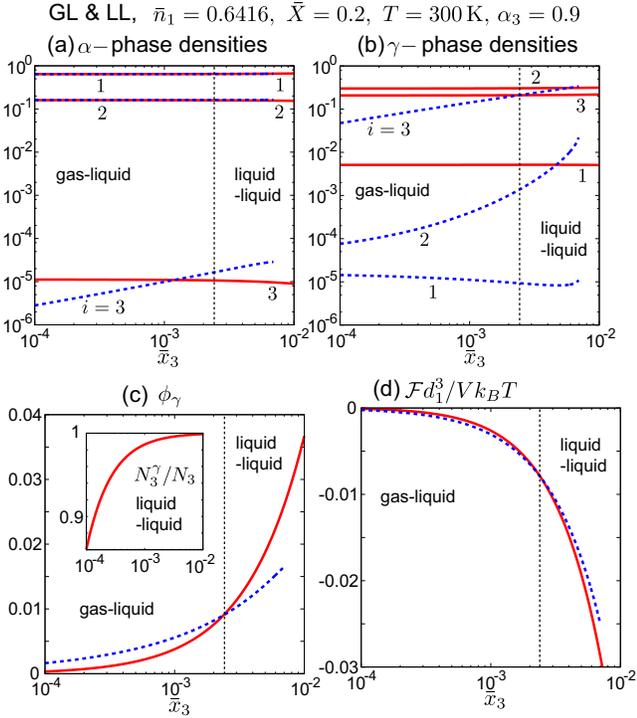}
\caption{Gas-liquid and liquid-liquid  coexistence 
with varying ${\bar x}_3=N_3/(N_1+N_2)$ at $T=300$ K, where 
 particle numbers $N_i={\bar n}_iV $ and  volume $V$ are fixed 
with  ${\bar X}=N_2/(N_1+N_2)=0.2$ 
and   ${\bar n}_1d_1^3=0.6416$  as in Fig.8(a). Here, 
 $\alpha_3=0.9$,  $\epsilon_{13}=\epsilon_{33}=0$, and $ 
\epsilon_{23}/k_B=300$ K. 
 Plotted are 
 (a)  $n_i^\alpha d_1^3$ ($i=1,2,3$)  
in the  $\alpha$ phase on the  gas-liquid branch (dotted lines) 
and on  the liquid-liquid branch (bold lines), 
 (b)  $n_i^\gamma d_1^3$  in the  $\gamma$ phase,  
  (c) volume fraction $\phi_\gamma$ 
of  the $\gamma$ phase,  and (d) 
free energy change ${ \cal  F}$ in Eq.(68) 
multiplied by $d_1^3 /V k_BT$. 
In (c), solute number $N_3^\gamma=V\phi_\gamma n_3^\gamma $ 
in the $\gamma$ phase 
divided by $N_3$ is also plotted (inset), which tends to 1 
with increasing ${\bar x}_3$.  In (d), difference of $\cal F$ 
between the two phases is very small.  
}
\end{center}
\end{figure}

\begin{figure}[t]
\begin{center}
\includegraphics[width=244pt]{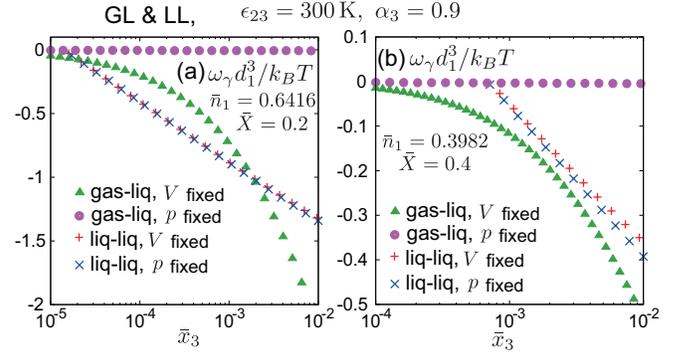}
\caption{
 Normalized grand potential  density 
 $\omega_\gamma d_1^3/k_BT$ in the $\gamma$ phase vs 
${\bar x}_3=N_3/(N_1+N_2)$  at $T=300$ K, where 
 ${\bar X}=N_2/(N_1+N_2)$ is (a) 0.2  and (b) 0.4. 
On  the  gas-liquid branch, the pressure increases and 
 $\omega_\gamma$ decreases with   
increasing ${\bar x}_3$  at fixed $V$  ({green filled triangle}), 
while $\omega_\gamma$  is nearly zero  at fixed  
 $p$  (purple filled circle).
On  the liquid-liquid branch, 
$\omega_\gamma$ is nearly the same for 
 fixed $V$  ({red} {$+$}) and  
 $p$  ({blue} {$\times$}).  
At fixed $V$ in (a), the data are common to those in Fig.9.
 Here,  $\alpha_3=0.9$,  $\epsilon_{13}=\epsilon_{33}=0$, and $ 
\epsilon_{23}/k_B=300$ K. }
\end{center}
\end{figure}

\begin{figure}[t]
\begin{center}
\includegraphics[width=240pt]{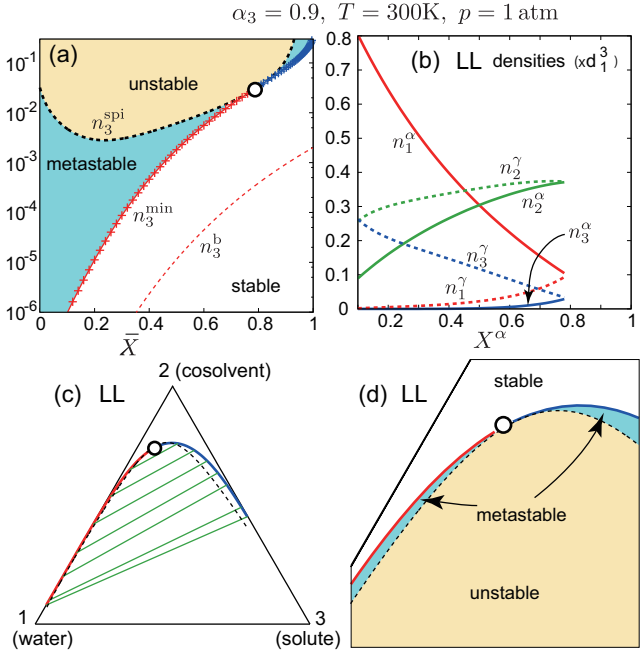}
\caption{ Phase  behaviors at $T=300$ K and 
$p=1$ atm  for   $\alpha_3=0.9$,  
$\epsilon_{13}=\epsilon_{33}=0$, and $ 
\epsilon_{23}/k_B=300$ K. 
(a) Phase diagram in the ${\bar X}$-$n_3 d_1^3$ plane 
on a semi-logarithmic scale.  Line 
$ n_3^{\rm min}$ vs $\bar X$  
separates stable and  metastable (blue) regions 
of liquid-liquid transition, on which  ${\bar X}= X^\alpha$. Unstable region is above   line $n_3^{\rm spi}$ 
 (yellow). Liquid-liquid critical point 
is marked by $\circ$. Binodal  line  is  in red  (blue) 
in the left (right) of  the critical point. 
Liquid-liquid phase separation   occurs   
above its binodal line  $ n_3^{\rm min}$ (in blue and yellow). 
Line $n_3^{\rm b}$ represents 
bubble  line at $p=1$ atm 
for gas-liquid transition in Eq.(66).   
(b) Coexisting densities   $n_i^\alpha$ 
and $n_i^\gamma$  vs $X^\alpha=n_2^\alpha/(n_1^\alpha+n_2^\alpha)$ 
in units of $d_1^{-3}$, which coincide at the   critical point. 
(c) Triangular phase diagram, 
where tie-lines (in green) connect coexisting two states. 
In (d), it is expanded near the critical point, where the 
metastable region (in green) is between the 
binodal and spinodal lines. In (a), (c), and (d), 
 dotted lines represent  spinodal. 
    }
\end{center}
\end{figure}
\begin{figure}[t]
\begin{center}
\includegraphics[width=200pt]{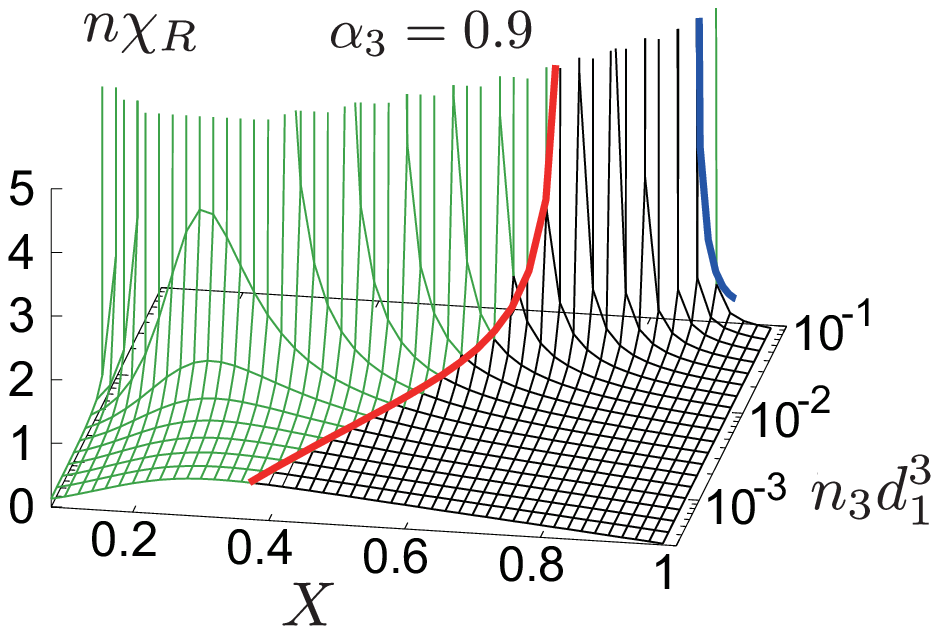}
\caption{ Variance of the concentration fluctuations 
 $n\chi_R$ ($n=n_1+n_2$) 
in Eq.(41) in one-phase states  at $T=300$ K 
 in the $X$-$n_3 d_1^3$ plane 
on a semi-logarithmic scale, where 
  $\alpha_3=0.9$,   $\epsilon_{13}=\epsilon_{33}=0$, and $ 
\epsilon_{23}/k_B=300$ K.  Contribution from  phase-separated 
droplets is not included. It grows on approaching  spinodal line 
and  critical point (see Fig.11). Stable region (right) 
and  metastable region (left)  are 
separated by liquid-liquid binodal  lines on the surface.  
}
\end{center}
\end{figure}

\subsection{Phase transitions at fixed $V$-$N_i$-$T$}

We  also examine  phase separation at fixed particle 
numbers $N_i$ in a fixed  volume $V$, where 
the initial one-phase state with densities 
$ {\bar n}_i= N_i/V $ is   metastable or unstable.  For  
  $\alpha_3=0.9$ and $\epsilon_{23}/k_B=300$ K, we then encounter 
both  gas-liquid and liquid-liquid phase 
transitions. The latter  was also found   
  for $\alpha_3=1.1$ and not for $\alpha_3=1.3$ 
if   the other  parameters were  unchanged. 
This   liquid-liquid  phase separation  
    occurs in  a wider parameter region 
for larger $\epsilon_{23}$, but   disappears   as $\epsilon_{23} \to 0$.

At  fixed $N_i$ and $V$, we consider  the  Helmholtz free energy. 
Its  change after phase separation is 
written  as 
\be 
 {\cal F} =
 V[ \phi_\gamma f_\gamma+ (1-\phi_\gamma) f_\alpha- {\bar f}], 
\en  
where    $\phi_\gamma$ is  the volume fraction of 
the $\gamma$ phase,  $f_\alpha$ and $f_\gamma$ 
are the values of $f$ in Eq.(52) in the two phases,   
 and ${\bar f}$ is the initial value of $f$. 
We  seek minima 
of  $\cal F$ imposing  the conditions  
$(1-\phi_\gamma) n_i^\alpha + \phi_\gamma n_i^\gamma= 
{\bar n}_i$ ($i=1,2,3$). To this end, we  
change $\phi_\gamma \to \phi_\gamma+\delta\phi_\gamma$ 
and $n_i^\gamma\to n_i^\gamma+\delta 
n_i^\gamma$    infinitesimally  to obtain  
the incremental change in 
 $\cal  F$ as 
\be 
\delta{\cal  F} 
= \sum_i (\mu_i^\alpha- \mu_i^\gamma) \delta N_i^\gamma+
 V(p_\alpha-p_\gamma) \delta\phi_\gamma, 
\en
where ($p_\alpha$, $\mu_i^\alpha$) and ($p_\gamma$, $\mu_i^\gamma$) 
are the values of $(p,\mu_i)$ in the two phases and 
$ N_i^\gamma= V \phi_\gamma n_i^\gamma$ are the particle numbers 
in the $\gamma$ phase. Thus, from   $\p {\cal  F}/\p \phi_\gamma
= \p {\cal  F}/\p n_i^\gamma=0$, we  
obtain  common values of $p$ and $\mu_i$ in the two phases.

Figure 8 displays $p_{\rm cx}$ 
  vs  ${\bar x}_3=N_3/(N_1+N_2)$  
at ${\bar X}= N_2/(N_1+N_2)=0.2$ 
 in the two phase transitions. 
We set  ${\bar n}_1=N_1/V$ equal to 
 (a) $0.6416 d_1^{-3}$ and 
(b) $0.6327 d_1^{-3}$, for which  the  initial 
pressure ${\bar p}_0=p({\bar n}_1,  {\bar n}_2, 0)$ 
 (without solute)  is (a) $1$ and (b) $-613.5$ atm. 
In (a), at ${\bar x}_3 =10^{-4}$,
$p_{\rm cx}$ is still  90 atm due to the presence of a gas fraction 
($\sim  10^{-3}$ in Fig.9(c))) 
on  the gas-liquid  branch but is close to ${\bar p}_0= 1$ atm 
on the liquid-liquid branch. 
In  (b), as ${\bar x}_3$ is decreased,   ${p}_{\rm cx}$ 
tends to $ p_{\rm cx}^0=0.075$ atm 
 on the gas-liquid branch  and to ${\bar p}_0=  -613.5$ atm 
on the liquid-liquid branch. Thus, in our model fluid, 
we can realize equilibrium liquid-liquid coexistence at 
 negative pressures\cite{Schneider}. 
Here,    $p_{\rm cx}$ depends    on   $n_3^\alpha$ 
sensitively because $\zeta_3$ in Eq.(6) 
is  large $(\sim 40$) in   liquid as in  Fig.2(c). 

 In  Fig.9, we further  show  
(a)  $n_i^\alpha$, (b) $n_i^\gamma$, 
(c)  $\phi_\gamma$,    and (d) $\cal F$  vs ${\bar x}_3$ for the case  
 of  Fig.8(a).   The  gas-liquid  branch 
 exists in the    range  $4\times 10^{-8}<
 n_3^\gamma d_1^3<0.337$ here, on which the 
$\gamma$ phase consists mostly of the solute. 
 On  the liquid-liquid branch,    $n_2^\gamma d_1^3 
(\sim 0.3)$ and  $ n_3^\gamma d_1^3 (\sim 0.2)$ are 
nearly constant (both increasing  by $3\%$ at ${\bar x}_3=10^{-2}$),  
while  $n_1^\gamma d_1^3(\sim 0.005)$ is  small. 
In (c),  $\phi_\gamma$ shrinks  
as ${\bar x}_3\to 0$. In  (d), $\cal F$ is slightly 
lower (higher) on  the gas-liquid branch 
than on the  liquid-liquid branch in the left (right) of the 
vertical  line at ${\bar x}_3\sim 2.5\times 10^{-3}$.  
However, from  $\cal F$, 
we cannot decide which transition occurs  in real situations.

To examine the metastability, we    introduce 
the grand potential density $ \omega(n_1,n_2,n_3)$  
by\cite{nano,Onukibook,Blander}  
\be
\omega = f-\sum_i {\bar \mu}_i n_i +{\bar p}
=\sum_i (\mu_i-{\bar \mu}_i) n_i +{\bar p}- p,
\en
where ${\bar \mu}_i$ and $\bar p$ 
are the  values of $\mu_i$ and $ p$ in the reference 
state with   $n_i={\bar n}_i$. Then, $\cal F$ is rewritten as 
\be
{\cal F} = V[ \phi_\gamma \omega_\gamma+ (1-\phi_\gamma) \omega_\alpha], 
\en  
where  $\omega_\alpha$ and $\omega_\gamma$  are 
the values of $\omega$  in the two phases.
Here,  $\omega_\alpha$ is of the second order in the deviations 
$n_i^\alpha- {\bar n}_i (\propto \phi_\gamma)$ 
as $\delta f_{\rm in}$ in Eq.(26), so $\omega_\alpha 
\propto \phi_\gamma^2$ as $\phi_\gamma \to 0$. On the other hand, 
we have  $\omega_\gamma<0$  when  
the the initial state with $n_i={\bar n}_i$ is metastable or unstable. 

In Fig.10, we plot $\omega_\gamma$ 
vs ${\bar x}_3$ at  $X^\alpha=$0.2 and 0.4.  For curves  at fixed $V$ 
we used the data in Figs.8(a) and 9, on which 
$\omega_\gamma<0$. On the gas-liquid branch, 
 $\omega_\gamma \to 0$ as 
$\phi_\gamma\to 0$, since ${\bar p}_0= 
p({\bar n}_1, {\bar n}_2, 0)=1$ atm 
is  close to $p_{\rm cx}^0$. 
The liquid-liquid branch  
exists only  for ${\bar x}_3>x_3^{\rm min}$, 
where  $\phi_\gamma \to 0$ 
and ${\bar n}_i \to  n_i^\alpha$ as  ${\bar x}_3\to 
x_3^{\rm min}$.
In Fig.10,   $x_3^{\rm min}$ is 
$1.41\times 
10^{-5}$ at  ${\bar X}=0.2$ 
and  $ 7.20\times 10^{-4}$ at  
${\bar X}=0.4$.

\subsection{Phase transitions at fixed $p$-$N_i$-$T$ } 

We next seek    two-phase coexistence 
in the isobaric condition 
  requiring $\mu_i^\alpha=\mu_i^\gamma$ 
and $p_\alpha= p_\gamma=1$ atm. 
The average densities ${\bar n}_i$ 
are chosen to satisfy  ${\bar p}= 
p({\bar n}_1, {\bar n}_2, {\bar n}_3)=1$ atm. 
This is needed since $p_\alpha \to {\bar p}$  
as $\phi_\gamma\to 0$. 
 If we increase  ${\bar n}_3$ gradually, 
${\bar n}_1$ and ${\bar n}_2$  decrease according 
to  Eq.(12) for each fixed ${\bar X}
= {\bar n}_2/({\bar n}_1+ {\bar n}_2)$. 
In this method,  we  vary  the volumes of the two phases, 
 $V_\alpha$ and $V_\gamma$, 
at fixed total particle numbers $N_i={\bar n_i}{\bar V}$, 
where $\bar V$ is the initial volume. 
Using  $\omega$ in Eq.(70), we write 
 the change in the Gibbs free energy after phase separation  as    
\be 
{\cal  G}= V_\alpha \omega_\alpha+ V_\gamma \omega_\gamma .
\en 
The  particle numbers $N_i^\alpha= V_\alpha n_i^\alpha$ 
and $N_i^\gamma= V_\gamma n_i^\gamma$  
in the two phases satisfy  $N_i^\alpha + N_i^\gamma  
= N_i$. For infinitesimal changes in these quantities, $\cal  G$ changes by 
\be 
\delta{\cal G}
= \sum_i (\mu_i^\gamma-\mu_i^\alpha)\delta N_i^\gamma
 + ({\bar p}- p_\alpha)\delta 
V_\alpha+({\bar p}- p_\gamma)\delta V_\gamma .
\en 
Thus, from  $\p{\cal G}/\p n_i^\gamma= \p {\cal G}/\p V_\gamma
= \p {\cal G}/\p V_\alpha=0$, we obtain   $\mu_i^\alpha=\mu_i^\gamma$ 
and $p_\alpha= p_\gamma={\bar p}$.    
On the gas-liquid branch, we have  
$V_\alpha\cong {\bar V}$. 
On the liquid-liquid branch, 
 $n_i^\gamma$  are nearly constant satisfying 
the space-filling relation  in Eq.(12), 
which are  slightly smaller  
than those at fiixed $V$.

In Fig.10, we additionally plot curves of 
$\omega_\gamma$ vs ${\bar x}_3$ for ${\bar X}=0.2$ and 0.4 
at fixed $p$ as well as those at fixed $V$. 
 On the gas-liquid branch at fixed $p$, $\omega_\gamma$ 
 nearly vanishes at any ${\bar x}_3$,  because 
$n_3^\gamma $ remains small ($\sim 6\times 10^{-4}d_1^{-3}$). 
On  the liquid-liquid branch, 
the two curves  of $\omega_\gamma$ at fixed $V$ and  $p$
almost  coincide and depend  on ${\bar x}_3$ 
logarithmically. These are because    $n_3^\gamma$ 
is changed only by a small amount 
($\ls 3\%$   in the figure) 
and the term  $-{\bar \mu}_3n_3^\gamma $  
in $\omega_\gamma$  gives rise to the 
contribution  $- k_BTn_3^\gamma
\ln {\bar x}_3 $ (see Eq.(80)).  

\subsection{Phase  diagrams at fixed $p$-$T$}

In  Fig.11,  the phase  behaviors are illustrated 
for $\alpha_3=0.9$ at  $T=300$ K and $p=1$ atm. 
 In (a), stable and    metastable  
   regions of the liquid-liquid phase transition  are separated by 
the binodal line  $ n_3^\alpha=n_3^{\rm min}$ in the 
${\bar X}$-$n_3 $ plane. Here,  $n_3^{\rm min}$ is  
the minimum of ${\bar n}_3= {\bar x}_3({\bar n}_1+ {\bar n}_2)$ 
on the isobaric  liquid-liquid branch with  
${\bar p}=p({\bar n}_1, {\bar n}_2, {\bar n}_3) =1$ atm (see Fig.10). 
At ${\bar n}_3= n_3^{\rm min}$, we have 
$\phi_\gamma = 0$,   ${\bar n}_i= n_i^\alpha$, and    
 ${\bar X}= X^\alpha$ and obtain $n_i^\gamma$ also.  
Above  the spinodal  line 
  ${\bar n}_3> n_3^{\rm spi}$, the one-phase states are unstable 
against  either of the two transitions. 
In (a), we also plot the bubble  line  ${\bar n}_3
= n_3^{\rm b}(\cong 0.8\times 
10^{-3}d_1^{-3}L_0$ here)   in Eq.(66). On the other hand, 
in the region  $n_3^{\rm b}< {\bar n}_3< n_3^{\rm spi}$,  
the one-phase states are   metastable 
with respect to the gas-liquid transition. 
In (b), we plot  the  densities $n_i^\alpha$ 
and $n_i^\gamma$ vs $X^\alpha$ for  the liquid-liquid  
transition, where  we have $n_i^\alpha= n_i^\gamma$ 
at a  critical point  given by   
 $X^\alpha=0.78$ and $n_3^\alpha=0.03 d_1^{-3}$ at fixed $T$ and $p$.  
In (c) and (d), we present the triangular phase diagram 
 of the liquid-liquid  
transition  using the molar fractions $x_i$. For $x_2 \ls 0.7$, 
 the binodal line is  very close to the triangle 
boundaries for the present   hydrophobic solute.

In  liquid-liquid coexistence, 
  the differential relation 
$\sum_i (n_i^\alpha- n_i^\gamma) d\mu_i^\alpha=0$ holds,  
where    $\mu_i^{\alpha}$  are the  $\alpha$-phase chemical potentials  
with  $\mu_3^{\alpha} \cong k_BT[
\ln (n_3^{\alpha}\lambda_3^3)+\nu_3(n_1^\alpha,n_2^\alpha)]$ 
from  Eq.(5). 
 We treat  the binodal density $ n_3^{\rm min}=n_3^\alpha$ 
as a function of $X^\alpha$ at fixed $T$ and $p$  to   find    
\be 
\bigg( \frac{\p \ln n_3^{\rm min}}{\p X^\alpha}  \bigg)_{T,p}= 
- g_3 + \frac{ {n_2^\gamma}{n_1^\alpha} - 
{n_1^\gamma}{n_2^\alpha}}{( n_3^\alpha-n_3^\gamma) n_\alpha^2 \chi} 
\quad ({\rm LL}), 
\en
using    Eqs.(30) and (A13). Here, 
 $g_3$ and $\chi$ are  the $\alpha$-phase values  and 
$n_\alpha= n_1^\alpha+ n_2^\alpha$. In Fig.11(a),  the  
 above derivative   is close  to  $-g_3$, where   
 the second term  in  
 Eq.(74) is  about  $0.1 g_3$ 
 and the critical value of  $g_3$ 
is   $-10.4$.

In  Fig.12, we plot the mean-field concentration variance 
 $\chi_R$ in Eq.(41) 
 in one-phase states in  the stable and metastable regions, where 
  $\alpha_3=0.9$,   $T=300$ K,  and $p=1$ atm.   
This   $\chi_R$  does not 
include the  contribution from  phase-separated  domains, 
but increases  above $\chi$ due to  
the solute-concentration  coupling in Eq.(29).  
We can approach  the critical point 
in Fig.11 passing through  the stable  region 
with respect to  the liquid-liquid 
transition   to encounter   the critical  scattering. 

Previously, Moriyoshi {\it et al.}\cite{Moriyoshi} 
found   a solute-induced 
critical point  in   a mixture of 
water, ethanol, and C8-alkanol. 
Recently, Anisimov's group\cite{Ani3}  measured 
 the  surface tension near a critical point in a mixture of   water, 
TBA, and cyclohexane, where the correlation length 
increased up to 9.5 nm.  Singular  behaviors  around such a  
unique critical point are  of great interest.

Though constructing a theory  of surfactant-free microemulsions is a future 
project, we make some comments. 
If the second species is   amphiphilic,  emulsification  can  occur in 
 the metastable region $n_3^{\rm min}<{\bar n}_3< n_3^{\rm spi}$, while 
 long-wavelength  fluctuations  
grow  up to   macroscopic sizes in the unstable region 
${\bar n}_3>n_3^{\rm spi}$.  (i) To support  this simple scenario, 
 Vitale and Katz\cite{Katz}  and  
Sitnikova {\it et al.}\cite{Wegdam} 
observed   micrometer-sized Ouzo droplets 
in a narrow metastable region and macroscopic phase separation 
in an  unstable region in  the phase diagram for 
strongly hydrophobic solutes   in   water-ethanol 
solvents.  This  was 
displayed  in a plane of the ethanol fraction 
and the logarithm of the  solute 
fraction as in    Fig.11(a). (ii) 
   Anisimov's  group\cite{Ani1,Ani2}  
 observed   droplets   with sizes of order  $100$ nm 
in a narrow region  in the triangular phase 
diagram  at small cyclohexane fractions 
 {\it outside the  binodal}    in a water-TBA solvent. 
They also  observed mesoscopic  droplets  
in wider regions outside the binodal 
for less hydrophobic solutes in the same solvent\cite{Ani2}. 
(iii) Through  the Ouzo process, 
mesoscopic  articles with sizes  of order 
100 nm can  be produced 
for various hydrophobic solutes including  polymers\cite{Aubry,Roger}    
  in water-hydrotrope mixtures.

\subsection{ Solute-induced nucleation  }

We now discuss 
 homogeneous nucleation\cite{Ox,Ox1,Langer,Blander,Onukibook} in 
a reference metastable  state with densities ${\bar n}_i$ 
and pressure $\bar p$. 
We suppose  a  spherical domain  with radius $R$  
in the limit $\phi_\gamma \to 0$. The  droplet  free energy  is  
given by  
\be 
 F(R)= (4\pi/3) R^3\omega_\gamma+ 4\pi R^2 \sigma ,
\en  
 where  $\omega_\gamma$ is defined in  Eq.(70)  
and the second term is the surface free energy. 
The   surface tension $\sigma$ is assumed to be a constant.  
This $F(R) $ depends on $R$,  $n_i^\gamma$,  and ${\bar n}_i$. 
For small changes in $R$ and $n_i^\gamma$ at fixed ${\bar n}_i$, 
$F(R)$ changes as 
\be 
\frac{\delta F(R)}{4\pi R }=  
 (R \omega_\gamma+ 2{\sigma})\delta R     
+\frac{R^2}{3}  \sum_i (\mu_i^\gamma-{\bar\mu}_i)\delta n_i^\gamma. 
\en 
We determine  $n_i^\gamma$  setting $\mu_i^\gamma= {\bar \mu}_i$ 
to seek a critical droplet. 
Then, Eq.(70) gives a simple relation, 
\be 
 \omega_\gamma=  {\bar p}- p_\gamma<0.
\en 
Since the second term  in Eq.(76) vanishes, 
 $F(R)$ has a { maximum} $F_c= F(R_c)$  at a critical radius $R_c$, where  
\be  
R_c= 2\sigma/(p_\gamma- {\bar p}),\quad 
F_c = (4\pi/3)\sigma R_c^2. 
\en 
These relations are general. In particular,  
we find  $\omega_\gamma\cong 
(1-n_1^\gamma/n_1^\alpha)({\bar p}-p_{\rm cx}^0)$ 
 for  weak metastability in 
 one-component fluids\cite{Langer,Onukibook},  where 
the condition $\mu_1^\gamma= {\bar \mu}_1$ 
becomes  $(p_\gamma-p_{\rm cx}^0)/ n_1^\gamma \cong 
 ({\bar p}- p_{\rm cx}^0)/n_1^\alpha$.

At the   gas-liquid transition 
with $n_3^\gamma \ll d_1^{-3}$,  
we find   $n_3^\gamma\cong  L_0^{-1}{\bar n}_3$ 
from $\mu_3^\gamma= {\bar \mu}_3$, so use of  
 $p_{\rm b}$ in   Eq.(67) yields 
\be 
\omega_\gamma \cong {\bar p}-  p_{\rm b}  \quad ({\rm GL}). 
\en 
At the  liquid-liquid transition, we use   
the  relation  $d\omega_\gamma \cong  \sum_{i} 
( { n}_i^\alpha- n_i^\gamma)
d{\bar \mu}_i =  ({ n}_3^\alpha- n_3^\gamma) d({\bar \mu}_3- \mu_3^{\rm min})$ 
with  $\mu_3^{\rm min}= \mu_3({\bar n}_1, {\bar n}_2, n_3^{\rm min})$. 
Here,  $n_3^{\rm min}({\bar n}_1, {\bar n}_2)$ is  the 
binodal solute density for  densities ${\bar n}_1$ and $ {\bar n}_2$.  
Thus,   
\be 
\omega_\gamma\cong -k_BT (n_3^\gamma-n_3^\alpha) 
 \ln ({\bar n}_3/ n_3^{\rm min})\quad 
({\rm LL}) ,
\en 
which is negative 
for ${\bar n}_3> n_3^{\rm min}$.  The curves of $\omega_\gamma$ 
at fixed $p$ in Fig.10 can  well be fitted to Eq.(80).  

\begin{figure}[t]
\begin{center}
\includegraphics[width=240pt]{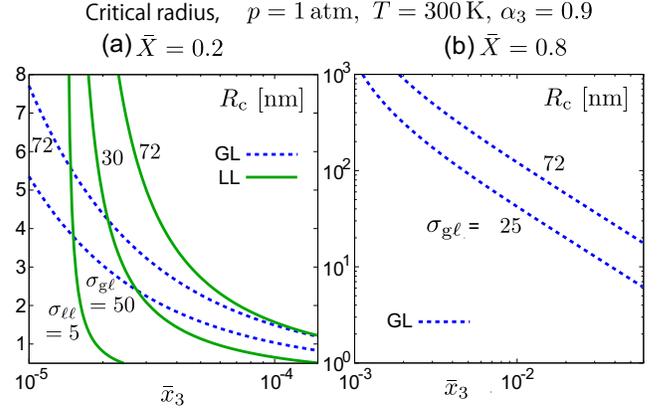}
\caption{
 Critical radius $R_c$  for  gas-liquid and  liquid-liquid transitions 
vs  ${\bar x}_3={\bar n}_3/({\bar n}_1+{\bar n}_2)$ 
 in metastable states with densities 
${\bar n}_i$   at $T=300$ K and $p=1$ atm. 
Here,   ${\bar X}={\bar n}_2/({\bar n}_1+{\bar n}_2)$ 
is (a) $0.2$ and (b) 0.8 
with   $\alpha_3=0.9$,   $\epsilon_{13}=\epsilon_{33}=0$, and $ 
\epsilon_{23}/k_B=300$ K.
In units of   mN$/$m, 
the gas-liquid surface tension 
  $\sigma_{{\rm g}\ell}$ is  (a) 72 and 50   and 
 (b) 72 and 25, while the liquid-liquid surface tension 
 $\sigma_{\ell\ell}$ is $72$, $30$, and $5$ in (a).   
}
\end{center}
\end{figure}

In our problem,  we  should note that  $\sigma$  depends 
on the composition $X^\alpha$ and 
take  different values in  gas-liquid and  liquid-liquid phase transitions.   
Hence,  we write   $\sigma= \sigma_{{\rm g}\ell}$ for 
 the gas-liquid case   and  $\sigma= 
\sigma_{\ell\ell}$ for the liquid-liquid case. 
(i) For binary water-alcohol mixtures, 
$ \sigma_{{\rm g}\ell}$ is decreased  
 by  interfacial  adsorption of alcohol
\cite{sigma1,sigma3,sigma4,sigma5}.  
For example,  at $X=0.05$, it is  $32$  mN$/$m for ethanol 
and $25$  mN$/$m for TBA, which are considerably 
 below the pure-water value $\sigma_{\rm g\ell}^{\rm w}=72$  mN$/$m.  
(ii) For  ternary mixtures of water, alcohol, and a hydrophobic solute, 
 $\sigma_{\ell\ell}$ should be    decreased  more strongly 
since  the oil side contains  hydrophobic molecules\cite{Fiore}. 
  It was well below $1$ mN$/$m 
as the critical pint was approached\cite{Zemb1,Zembreview,Ani3}.

In Fig.13, we plot  $R_c$ in Eq.(77) 
vs  ${\bar x}_3$  for (a) ${\bar X}=0.2$ and (b) 0.8 
 in the  metastable states in Fig.11(a), where  
$\alpha_3=0.9$,  $T=300$ K,  and $p=1$ atm.  
We write   $R_c$  for a few plausible 
values of $\sigma_{{\rm g}\ell}$ 
and $\sigma_{\ell\ell}$.  
In (a), $R_c$ for the liquid-liquid transition 
is  well below 1 nm 
with increasing ${\bar x}_3$, where the nucleation barrier 
is negligible. In (b), 
$R_c$ for the gas-liquid transition 
remains in the range $R_c\gs 10$ nm even 
for  $\sigma_{{\rm g}\ell}=25$ mN$/$m  
 on approaching the critical point of the liquid-liquid transition, 
where  homogeneous bubble nucleation 
is  highly improbable. Note  that bubble nucleation 
in  pure water was detected    when   $R_c$ was  about  
 1 nm at large negative pressures at $T \sim 300$ K\cite{Angell,Caupin}. 
 
In   water-alcohol 
solvents,   nanometer-sized solute-rich aggregates 
preexist with  alcohol molecules covering them 
in one-phase states\cite{Zemb1,Zemb,Zemb2,Hori,Ani1}.    Then, they  
should   play  the role of  
{\it embryos} of critical  liquid droplets   
consisting of solute and alcohol molecules.  
This much amplifies the prefactor $I_0$ 
of the nucleation rate $I= I_0 \exp(-F_c/k_BT)$ of the 
liquid-liquid transition\cite{Langer,Onukibook}. 
However, we can well expect  occurrence of 
the gas-liquid  transition 
for weak  solute-cosolvent 
attraction (for smaller 
 $\epsilon_{23}$ in our scheme) 
 (see the first paragraph in Sec.IVB).

   Jin {\it et al.}\cite{S7} reported observations of 
 air nanobubles  in  mixtures 
of water and a neutral organic
water-soluble   species such as tetrahydrofuran, 
where the latter has an   amphiphilic  nature 
and   accumulates  at gas-liquid interfaces. 
These bubbles    were  removed by repeated filtration and
 regenerated by air injection. Note that the method of 
  gas injection   is widely  used 
to produce nanobubbles, where 
large bubbles are fragmented into small ones\cite{Ohgaki}.

\section{Summary and remarks}

We have studied   a continuum theory of solvation 
 and phase transitions in nonionic ternary mixtures 
 varying the solvent composition $X$ and the solute 
density $n_3$. 
Main results are summarized as follows.\\
(1) In Sec.II, we have  presented  a   theory 
of  ternary mixtures  
using the second virial  expansion of the free energy density $f$ 
with respect to  $n_3$. 
We have clarified relationship among the  solvation 
chemical potential $k_BT \nu_3$, the  partial volumes  
$v_i$, the Kirwood-Buff integrals $G_{ij}$ (and the dilute limits $G_{i3}^0$), 
 the solute spinodal density  $  n_3^{\rm spi}$, 
the second virial osmotic coefficient $B_2$, 
and the Gibbs transfer free energy $\Delta G$.  
 We have confirmed  the space-filling relation  
 $\sum_i v_i n_i=1$  for slowly varying densities $n_i$ 
 in the limit of small compressibility. In  
nearly incompressible mixtures, the  solute-concentration coupling 
 grows as   in  Fig.4 when the solute interacts with 
the two solvent species  differently. 
It  is much  amplified 
in water--alcohol mixtures 
in the presence of  micelle-like aggregates 
\cite{Zemb,Zemb1,Zemb2,Hori,Ani1}.\\
(2) In Sec.III, we have   investigated 
the solvation properties using  the    MCSL model. 
The parameters of our model have been chosen 
to realize the following at $T=300$ K. (i) The saturated vapor pressure 
 $p_{\rm cx}^0$ is very small ($\ll 1$ atm), 
(ii) the two solvent species are miscible 
in any  proportion at $p=1$ atm, and 
(iii) the solute interacts repulsively 
 with the first species but attractively with the second. 
In this situation, the solvent-mediated 
solute-solute interaction mainly arises from 
the solute-concentration coupling for not small $X$. 
 In the case of   one-component solvents ($X=0$), 
we have examined the  effective solute interaction 
vs the  diameter ratio $\alpha_3= d_3/d_1$ in Fig.5. \\
(3) In Sec.IV,  we have   examined  
solute-induced gas-liquid and liquid-liquid phase transitions.
We have determined the solute spinodal density, 
the metastability conditions, and 
the binodal  conditions. 
We have  presented  phase diagrams in Fig.11 
and  discussed homogeneous nucleation.\\

Finally, we give some  remarks. 
(1) We need to  calculate  the surface tension 
 for   amphiphilic cosolvents.  
There should be 
  significant   differences between methanol, 
ethanol,  and TBA   in the  formation of Ouzo domains.   
In the pre-Ouzo regime,  the hydrotrope  
molecules consist of those forming micells 
and those in monomeric states\cite{Tan,Zemb,Shimizu}. 
(2)    Kinetics of solute-induced  phase transitions  
  should  be  investigated, 
where slow solute diffusion is crucial\cite{nano,Wegdam,Roger}. 
(3) We can  add    a salt to water-hydrotrope mixtures\cite{Oka,Misawa}. 
It is of interest how  
an antagonistic salt (composed of 
hydrophilic  and hydrophobic ions) 
can induce aggregates in such  mixture solvents\cite{anta}, 
where an electric field can be applied\cite{Yabu}. 
(4) Microscopically,  the hydrogen-bonding 
structures in aqueous mixtures 
are  strongly deformed around  amphiphilic  and hydrophobic molecules. 
Thus,  inputs from molecular dynamics simulations 
are highly  informative\cite{Mochi,Lee,Zemb2,Hori,Ani1,Fiore,Dougan,Patey1,Bag}.

\begin{acknowledgments}
We  acknowledge support by JSPS KAKENHI Grant 
(No. JP18K03562 and No.15K05256).
 We would like to  thank K.  Koga and T.  Sumi for informative
discussions.  
\end{acknowledgments}

\vspace{4mm}
\noindent{\bf Appendix A:  General relations 
of partial volumes and  Kirkwood-Buff integrals }\\
\setcounter{equation}{0}
\renewcommand{\theequation}{A\arabic{equation}}

Let  a  nonionic fluid mixture 
be in  a macroscopic volume $V$ with 
 particle numbers   $N_i= n_i V$. 
The   partial volumes, written as ${\bar v}_{i} $ here, 
 are     defined by
\be 
 {\bar v}_{i} =
 ({\p V}/{\p N_i})_{T,p, \{N_{{j} \neq {i}}\}},  
\en  
where  $N_j$ with ${j}\neq {i}$ are fixed in  
 the derivative with respect to $N_i$. 
We obtain  
${\bar v}_i=  
({\p \mu_i}/{\p p})_{T,\{n_j\}}$ from 
$dG= -Vdp +\sum_i \mu_i dN_i$ at fixed $T$ 
for the Gibbs free energy $G$. 
Since  $V$ is an extensive quantity, 
we find the sum rule, 
\be
\sum_j {\bar v}_{j} n_j =1. 
\en 
The differential form of $V$ is then given by 
\be 
dV= \sum_j {\bar v}_{j} dN_j - \kappa V dp + \alpha_p V dT. 
\en 
Here, 
$\kappa=-(\p V/\p p)_{T,\{ N_j\}}/V$ is the isothermal compressibility 
and  $\alpha_p $ is the isobaric thermal expansion coefficient. 
In Eq.(A3) we set $N_j=Vn_j$  to  obtain  
\be 
 \sum_j {\bar v}_{j} dn_j - \kappa dp  +   \alpha_p  dT=0, 
\en 
with the aid of  Eq.(A2). 
Thus, we can express ${\bar v}_{i}$  as  
\bea
{\bar v}_{i}&=&\kappa (\p p/\p n_i)_{T,  \{n_{{j} \neq {i}}\}  }\\
&=& k_BT  \kappa \sum_j  n_j I^{ij} .  
\ena 
In deriving Eq.(A6)  we use  the Gibbs-Duhem relation 
($dp= \sum_j n_j d\mu_j$ at fixed $T$) and  define   
\be 
I^\ab=(k_BT)^{-1} (\p \mu_j/\p n_i)_{T,  \{n_{k\neq j}\}} .
\en
We obtain   Eqs.(10) and (11) from Eq.(A5). 

In the grand canonical ensemble,  we may  consider the fluctuations  
of  the particle numbers,  written as  $\delta {\hat N}_i$, 
in a region with  volume $V$. The    $I_{ij}$ in Eq.(15) 
can be related to the variances of  $\delta {\hat N}_i$  as     
\be 
I_{ij}= V^{-1}   \av{\delta {\hat N}_i \delta {\hat N}_j}=  
k_BT(\p n_i/\p \mu_j)_{T, \{\mu_{k\neq j}\}}  .
\en  
We notice that the matrix $I^\ab$ in Eq.(A7)  
is equal to  the inverse of the variance matrix $I_\ab$ in Eq.(A8).
Then,  
\bea  
&&
k_BT \kappa n_i= \sum_{j}  I_{ij} {\bar v}_j, \\
&&\hspace{-5mm} k_BT \kappa
= [ \sum_\bc  I^{\bc} n_j  n_k]^{-1}= \sum_\bc  I_{\bc} {\bar v}_j  {\bar v}_k .\ena 

It is easy to express $I^{ij}$ 
in terms of $I_{ij}$ for three component systems. 
Furthermore, for small $n_3$, $I_{ij}$ and $I^{ij}$ for 
the solvent components ($i,j=1,2)$ smoothly 
tend to those of $n_3=0$, while $I^{33}$ grows as $n_3^{-1}$ and  
\be
I^{i3}
 = -I^{i1}n_1 G_{13}-I^{i2}n_2 G_{23}+\cdots 
\quad (i=1,2). 
\en 
From  Eqs.(4), (7), (A7), and (A11) we find  
\bea 
\nu_{3i}&=& (k_BT)^{-1}\lim_{n_3 \to 0} 
(\p \mu_i/\p n_3)_{T, n_1, n_2}
\nonumber\\
&=& -\lim_{n_3 \to 0} [I^{i1}n_1 G_{13}+I^{i2}n_2 G_{23}].
\ena 
We then obtain Eqs.(17) and (31).
   
We finally  consider the composition $X$ for binary mixtures. 
Using  Eqs.(A6) and (24), we rewrite its   
differential relation $dX = (n_1 dn_2- n_2 dn_1)/n^2$ at fixed $T$  as    
\bea 
dX & = & (n^2\chi/k_BT)({\bar v}_1 d\mu_2- {\bar v}_2 d\mu_1)\nonumber\\
&=& (n\chi/k_BT)[({\bar v}_1-{\bar v}_2) dp+d\Delta].
\ena 
with  $\Delta= \mu_2-\mu_1$. For $dp=0$ we find  Eqs.(23) and (74).

\vspace{4mm}
\noindent{\bf Appendix B: Gas-liquid 
coexistence in  one-component solvent: 
Corrections to  $L$ and $k_{\rm H}$  }\\
\setcounter{equation}{0}
\renewcommand{\theequation}{B\arabic{equation}}

In a one-component solvent, 
 we examine  gas-liquid coexistence 
 with a  solute to linear 
order in its density. 
The  solvent and solute densities  are  
$n_1^\alpha$ and $n_3^\alpha$  in liquid  
 and   $n_1^\gamma$ and $n_3^\gamma$  in gas. 
Without solute, the  solvent 
density is  $n_\ell$ in liquid  and $n_g$ in gas. 
After the doping, it is changed by  
 $\delta n_1^\alpha= n_1^\alpha- n_\ell$ 
in liquid and  $\delta n_1^\gamma = n_1^\gamma- n_g$ in gas.

The deviations 
 $\delta\mu_1$,  $\delta\mu_3$, and $\delta p_{\rm cx}$ 
 assume common  values in the two phases. The 
Gibbs-Duhem relation gives 
\be 
\delta p_{\rm cx}= n_\ell\delta\mu_1+ k_BT n_3^\alpha 
= n_g\delta\mu_1+ k_BT n_3^\gamma, 
\en 
Here, for any physical quantity $\cal A$, we use  the symbol 
 $[{\cal A}]= {\cal A}_\alpha-{\cal A}_\gamma$,  
 where the values of $\cal A$ in the two phases are written 
with  $\alpha$ and $\gamma$. 
We then  solve Eq.(B1) as  
\bea 
&&\hspace{-10mm}
\delta\mu_1/k_BT  =  -  [n_3]/ [n_1], \\
&&\hspace{-1cm}\delta p_{\rm cx}/k_BT = 
 (n_1^\alpha n_3^\gamma-n_1^\gamma n_3^\alpha)/[n_1] \nonumber\\
&&\hspace{3mm}  = (n_\ell /L_0- n_g) n_3^\alpha/(n_\ell-n_g)
\ena
Thus, $\delta p_{\rm cx}$ is positive  (negative) 
 for $L_0 n_g/ n_\ell<1$ ($>1$). This occurs 
for $(\Delta G)_3/k_BT> -10.6$ ($<-10.6$)) 
for ambient water.  Far from the criticality, 
we have $\delta p_{\rm cx} \cong k_BT n_3^\gamma$ 
for  hydrophobic solutes as in  Eq.(66), 
but the shift is  small  for  hydrophilic solutes  
with $L_0\gg 1$.

From Eq.(4)  $\delta\mu_1$ is written as 
\be 
\delta \mu_1/k_BT 
= \delta n_1^\alpha/K_\ell + \nu_\ell n_3^\alpha
= \delta n_1^\gamma /K_g  + \nu_g n_3^\gamma,
\en
where $(K_\ell, \nu_\ell)$ 
and $(K_g,\nu_g)$ are the values of 
$K= n_1^2 k_BT \kappa$ and $\nu_{31}=\p \nu_3/\p n_1$ 
in liquid and gas, respectively, with $\kappa$ being the compressibility. 
Here,  Eq.(34) gives   $K\nu_{31}= n_1 v_3^{\rm in}$. 
From Eqs.(B2) and (B4) we find   
\bea
\delta n_1^\alpha &=& -K_\ell([n_3]/[n_1]+ \nu_\ell n_3^\alpha),\\
\delta n_1^\gamma &= &-K_g([n_3]/[n_1]+ \nu_g n_3^\gamma),
 \ena 
In gas, $K_g \cong n_g$ and $\nu_g =\lim_{n_i \to 0} (\p^2 p/\p n_1\p n_3)$.
Far from the criticality,  
$\delta n_1^\gamma$ is very small ($\sim  (n_g/n_\ell)n_3^\gamma$).

For the solute we use Eqs.(5) and (48) to obtain  
\be 
[\ln n_3] + \Delta G/k_BT=- [\nu_{31}\delta n_1] -[ U_{33} n_3]/k_BT,
\en 
where the left hand side is $\ln (L/L_0) $. 
With the aid of Eqs.(34) and (36), substitution of 
 Eqs.(B5) and (B6) into the right hand side  yields 
\be 
L/L_0-1 = [n_1 v_3^{\rm in}] [n_3]/[n_1]- 2[B_2 n_3]. 
\en
The correction to 
$k_{\rm H}$ 
 can  be calculated from  
\be 
\ln (k_{\rm H}/k_{\rm H}^0)=(\delta n_1^\alpha+ n_3^\alpha)/n_\ell+ 
\nu_\ell \delta n_1^\alpha + U_{33}^\alpha n_3^\alpha/k_BT ,
\en  
where the quantities in liquid appear and 
 the first term arises  from 
$\ln [(n_1^\alpha+ n_3^\alpha)/n_\ell]$. 
We thus find Eq.(51).

\begin{figure}[tbp]
\begin{center}
\includegraphics[width=235pt]{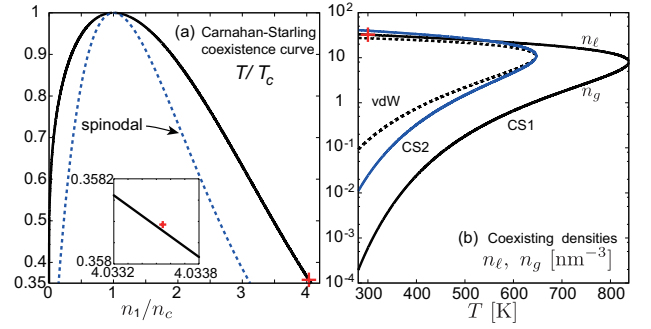}
\caption{ Phase diagram of  one-component  solvent 
  with 
  Carnahan-Starling  and  van der Waals interactions. 
(a) Coexistence and spinodal curves 
 in the $T/T_c$-$n_1/n_c$ plane. 
Reference state in this paper is at $T=300$ K  slightly above  
the coexistence curve  (red {$+$}) (see inset). 
(b) Coexisting liquid and gas densities 
$n_\ell$ and $n_g$ vs  $T$. On curve CS1 (bold line), 
use is made of  Eq.(60), for which 
  $p_{\rm cx}^0=0.031$ atm  at $T=300$ K.  
Curves of  CS2 (bold line) 
 and vdW (dotted line) are obtained from the critical-point 
fitting. 
}
\end{center}
\end{figure}

\vspace{4mm}
\noindent{\bf Appendix C: Phase diagram from CS  model}\\
\setcounter{equation}{0}
\renewcommand{\theequation}{C\arabic{equation}}

For   one-component fluids with density 
$n_1$,   Fig.14(a) displays 
 a phase diagram, 
where  the  steric  part is   given by 
 the Carnahan-Starling (CS)  model  and 
 the attractive part  by 
the van der Waals model\cite{CS1972,Higgens,Sadus} with  
   $d_1$ and $w_{11}$ being  independent of $T$ 
and $p$.  The pressure is given by    
\be 
p= { k_BTn_1} \bigg[  1+ \frac{4\eta_1-2\eta_1^2}{(1-\eta_1)^3} 
\bigg] - \frac{1}{2}{w_{11}} n_1^2 , 
\en 
where  $\eta_1=\pi d_1^3n_1/6$.
The  critical temperature $T_c$,  
density $n_c$, and   pressure $p_c$ are  calculated 
 as\cite{CS1972,Sadus} 
\bea
&& \hspace{-1cm} 
 k_BT_c =9.01 \times 10^{-2}  w_{11}d_1^{-3}, 
\quad n_c =0.249 d_1^{-3},\nonumber\\
&& \hspace{-5mm} 
 p_c/n_c k_BT_c=0.359. 
\ena

In this paper,  $d_1$ and $w_{11}$ are  given by 
 Eq.(62), which yield the coexistence  pressure  
 $p_{\rm cx}^0=0.031$ atm of water at $T=300$ K  (CS1 fitting). 
This gives   $T_c=838$ K and  $ n_c =0.247d_1^{-3}= 9.1$ nm$^{-3}$, 
which are  not far from   the water values    
  ($T_c=647.1$ K and $n_c=10.8$ nm$^{-3}$). 
 We can  also determine $d_1$ and $w_{11}$ 
to obtain $T_c$ and $n_c$ of  water (CS2 fitting), 
for which  $p_{\rm cx}^0=0.72$ atm at $T=300$ K. 
In Fig.14(b), curves CS1 and CS2 give  
the coexisting densities $n_\ell$ and $n_g$ 
vs $T$ from these  fittings, which are compared  
with curve vdW  from  the critical-point fitting of 
the van der Waals model. In  the  CS  model,  the 
 gas density  $n_g (\cong  p_{\rm cx}^0/k_BT)$ 
can be very small far  below $T_c$.

\vspace{4mm}
\noindent{\bf Appendix D: Summary of  MCSL   model}\\
\setcounter{equation}{0}
\renewcommand{\theequation}{D\arabic{equation}}

We     summarize  the multi-component 
MCSL   model \cite{Man}. Using   $d_i$, 
we write the hard-sphere  volume fractions  as 
$\eta_i={\pi} n_i d_i^3/6$ 
 and  the total one as   $\eta = \sum_j \eta_j$.  
The hard-sphere  free energy density 
$f_h$ in Eq.(52) can be simply  expressed in terms of 
$u=\eta/(1-\eta)$  as 
\bea 
&&\hspace{-1cm}
{f_h}/{k_B Tn} = 4u+u^2 -{3}y_1u-{3}(y_1+y_2){u^2}/2 \nonumber\\
&+&(y_3-1)[u-{u^2}/{2}- \ln (1+u)], 
\ena 
where   $n=\sum_j n_j$. We  define 
 $y_1$, $y_2$, and $y_3$  as    
\bea 
&&\hspace{-13mm}
y_1 = \sum_{i>j}\frac{(d_i+d_j)\Delta_{ij}}{(d_i d_j)^{1/2}},\quad 
y_2=  \frac{\xi_2}{\eta} \sum_{i>j} \Delta_{ij} (d_i d_j)^{1/2} ,
\nonumber\\ 
&&\hspace{5mm}
y_3 = 6 \xi_2^3/{\pi \eta^2n},
\ena 
where      
$\Delta_{ij}=  ({\pi }/{6})(d_i-d_j)^2 (d_i d_j)^{1/2}
{ n_i n_j}/{n\eta}$, 
and     $\xi_\ell 
=(\pi/6)\sum_i n_i d_i^\ell$ ($\ell=1,2,3$).

We  then calculate 
the hard-sphere part of the  chemical potential $\mu_{{\rm h}i} =
\p f_h/\p n_i$ in Eq.(56)  as   
\bea 
\frac{\mu_{{\rm h}i}}{k_B T}&=&
(3\gamma_i^2- 2\gamma_i^3-1) \ln(1-\eta)
+ ( 2u^2+ 3u-2) u\gamma_i^3 
 \nonumber\\
&&\hspace{-0.7cm}
+\frac{3 \xi_2 d_i}{1-\eta}   \bigg(1+ \frac{\xi_1 d_i}{\xi_2}+ 
\frac{\xi_0d_i^2}{3\xi_2} + \frac{\xi_1d_i^2+\gamma_i}{1-\eta}\bigg ) ,
\ena 
where      $\gamma_i= \xi_2 d_i/\eta$. 
In ternary mixtures,   
$\mu_{{\rm h}3}$  yields   $\nu_3$ 
  and $U_{33}$ as  $n_3\to 0$ from 
 Eqs.(57) and (58). They 
 have third-order and sixth-order  polynomial forms as\cite{Lada}  
\bea 
&&\hspace{-5mm}
\nu_3=-\ln(1-\eta) +D_1 \alpha_3+D_2 \alpha_3^2 +D_3 \alpha_3^3 ,\\  
&&n{U_{33}}= {k_BT} \sum_{0\le k\le 6} W_k \alpha_3^k,  
\ena 
where the  coefficients $D_k$ and $W_k$ 
depend on $n_1 $ and $n_2$.
Note that $n_3$ appears in the form of  $d_3^3 n_3$ 
in $f_h$ in Eq.(D1), yielding  the $\alpha_3^3$ term 
in $\nu_3$ and the $\alpha_3^6$ term in $U_{33}$ 
(see    Figs.2 and 5). 
In particular, we  express $D_3$ and $W_6$  as   
\bea 
&&\hspace{-1cm} 
D_3= B^3[ 2u^3+ u^2-2u -2\ln(1-\eta)] +  3 A_1 Bu^2/A_3\nonumber\\
&&\hspace{-0.8cm} +  u/A_3  -
(8\sqrt{2} /3)(\epsilon_{13} \eta_1+\epsilon_{23}\alpha_2^{-3}\eta_2)/k_BT, 
\\
&&\hspace{-1cm}
 W_6= (B^3/A_3)[ 5u^3+ 6u^2 -6\ln(1-\eta)] \nonumber\\
&&\hspace{-0.4cm} + (1+ 6 A_1Bu)u^2/A_3 ,
\ena  
where 
   $A_\ell=
 1+X(\alpha_2^\ell-1)$, 
$B=A_2/A_3$, $\eta=\eta_1+\eta_2$, 
 and $u= \eta/(1-\eta)$. The  $D_3$ and $W_6$ 
 are large for  $\eta\gs 0.5$.  
For example, we have   $D_3= 8.05- 1.98 \epsilon_{13}/k_BT$ 
 for $X=0$ and $\eta=0.526$ 
and $D_3=6.00-(1.27\epsilon_{13}+0.317\epsilon_{23})/k_BT$
for $X=0.2$,  $\eta=0.520$, and $\alpha_3=0.9$. 
Thus,  the coupling coefficients 
$\zeta_3$ in Eq.(8) and $g_3$ in Eq.(30) 
increase as $\alpha_3^3$ for $\alpha_3 \gs 1$, leading to 
  Eqs.(63) and (64).

\end{document}